\documentclass[10pt,journal]{IEEEtran}

\usepackage{amsmath,amsfonts}
\usepackage{txfonts}
\usepackage{booktabs}
\usepackage{capt-of}
\usepackage[nocompress]{cite}
\usepackage{graphicx}
\usepackage{url}
\usepackage[table]{xcolor}
\usepackage{array,multirow,tabularx}
\usepackage[colorlinks=true,linkcolor=blue,citecolor=black,urlcolor=black]{hyperref}

\usepackage{orcidlink}
\graphicspath{{figures/}}
\newcolumntype{D}{>{\centering\arraybackslash}X}
\newsavebox{\areabreakdownbox}

\newcommand{\tcadrunningheader}{IEEE TRANSACTIONS ON COMPUTER-AIDED DESIGN OF INTEGRATED CIRCUITS AND SYSTEMS,~VOL.~XX, NO.~XX, XX~2026}

\makeatletter
\def\ps@arxivfirstpage{%
  \ps@IEEEtitlepagestyle
  \def\@oddhead{\hbox{}\normalfont\scriptsize
    Preprint. Submitted to IEEE Transactions on Computer-Aided Design of Integrated Circuits and Systems.\hfil\thepage}%
  \let\@evenhead\@oddhead
}
\makeatother

\begin{document}

\title{\fontsize{28}{33}\selectfont MergeASR: Efficient Automatic Speech Recognition with Lightweight Cache-Based Token Merging}

\author{Wenhui Ou\,\orcidlink{0009-0009-6779-5841},
        Qiawei Zheng\,\orcidlink{0009-0009-6944-2548},
        Pingcheng Dong\,\orcidlink{0000-0003-3975-286X},
        Zhuoyu Wu\,\orcidlink{0000-0002-9075-7626},
        Dongjun Wu\,\orcidlink{0009-0007-1110-5798},
        Yijian Wu\,\orcidlink{0009-0002-2320-4826},\\
        Yu Sen To\,\orcidlink{0009-0000-6934-8337},
        Zheng Wang\,\orcidlink{0000-0003-2855-9570},~\IEEEmembership{Member,~IEEE},
        and C. Patrick Yue\,\orcidlink{0000-0002-0211-2394},
        ~\IEEEmembership{Fellow,~IEEE}%
\thanks{Wenhui Ou and Qiawei Zheng contributed equally to this work.
(Corresponding author: C. Patrick Yue.) This work was supported in part by
the Areas of Excellence Scheme of Hong Kong under Grant AoE/E-601/22-R and in
part by the General Research Fund of Hong Kong under Grant 16221425.}%
\thanks{Wenhui Ou, Qiawei Zheng, Pingcheng Dong, Dongjun Wu, Yijian Wu, Yu Sen To, and
C. Patrick Yue are with the Department of Electronic and Computer Engineering,
The Hong Kong University of Science and Technology, Clear Water Bay, Kowloon,
Hong Kong (e-mail: eepatrick@ust.hk).}%
\thanks{Zhuoyu Wu is with the School of Information Technology, Monash
University Malaysia, Jalan Lagoon Selatan, 47500 Bandar Sunway, Selangor Darul
Ehsan, Malaysia. Zheng Wang is with the School of Integrated Circuits,
Nanjing University, Suzhou, China.}}

\markboth{\tcadrunningheader}{\tcadrunningheader}
\IEEEpubid{1937-4151~\copyright~20XX~IEEE.}

\maketitle
\thispagestyle{arxivfirstpage}

\begin{abstract}
Dense acoustic tokens in automatic speech recognition (ASR) models lead to
substantial computational demand. Token merging, widely applied in vision
Transformers, is a promising solution that can reduce the number of tokens by
merging similar ones while preserving output accuracy. However, applying token
merging directly to ASR makes it hard to achieve a good tradeoff between
inference efficiency and transcription quality due to ASR's heterogeneous
encoder--decoder architecture and distinct acoustic-token similarity patterns.
In this paper, we propose MergeASR, a hardware--software co-designed
token-merging framework for efficient ASR inference. MergeASR adopts a
component-aware merging policy to avoid redundant matching in different ASR
components. Then, a cache-based token matching method is applied to efficiently
identify similar tokens by exploiting temporally local similarity patterns among
acoustic tokens. Finally, we design a lightweight, plug-and-play token-merging
engine that hides matching and merging latency during ASR execution with minimal
hardware overhead. Experiments demonstrate that, for real-time transcription,
MergeASR provides 1.49$\times$ inference speedup, with a slight increase of
only 0.38 percentage points in word error rate on challenging speech recordings.
\end{abstract}

\begin{IEEEkeywords}
Automatic Speech Recognition (ASR), Token Merging, Accelerator,
Hardware-Software Co-design
\end{IEEEkeywords}

\section{Introduction}
\label{sec:introduction}

\IEEEPARstart{A}{utomatic} speech recognition (ASR) converts spoken language
into text, providing a natural interface between people and intelligent
systems. Recently, ASR has become an essential front end for services built on
large language models (LLMs), translating spoken requests into the text
consumed by downstream language understanding and generation~\cite{cui2025recent}.
However, the inference cost of modern ASR models renders high-quality
transcription impractical on many edge devices~\cite{wang2025whisperflow} and
expensive to serve at scale in the cloud~\cite{kamahori2025liteasr}.

\begin{figure}[!t]
    \centering
    \includegraphics[width=0.85\columnwidth]{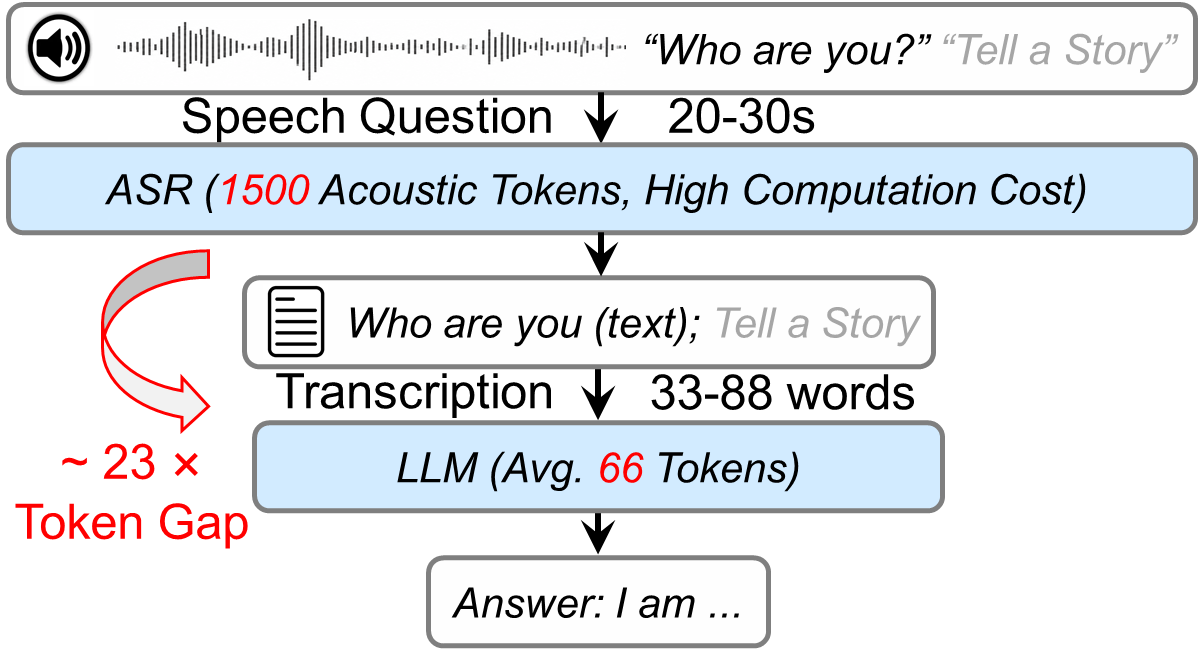}
    \caption{ASR models process far more tokens per spoken context than LLMs,
    leading to high computational demand. Evaluated on LibriSpeech using
    Whisper-Medium (0.76B)~\cite{radford2023robust} and
    Qwen2.5-0.5B~\cite{hui2024qwen2}.}
    \label{fig:asr-llm-token-imbalance}
\end{figure}

\begin{figure*}[!t]
    \centering
    \includegraphics[width=0.95\textwidth]{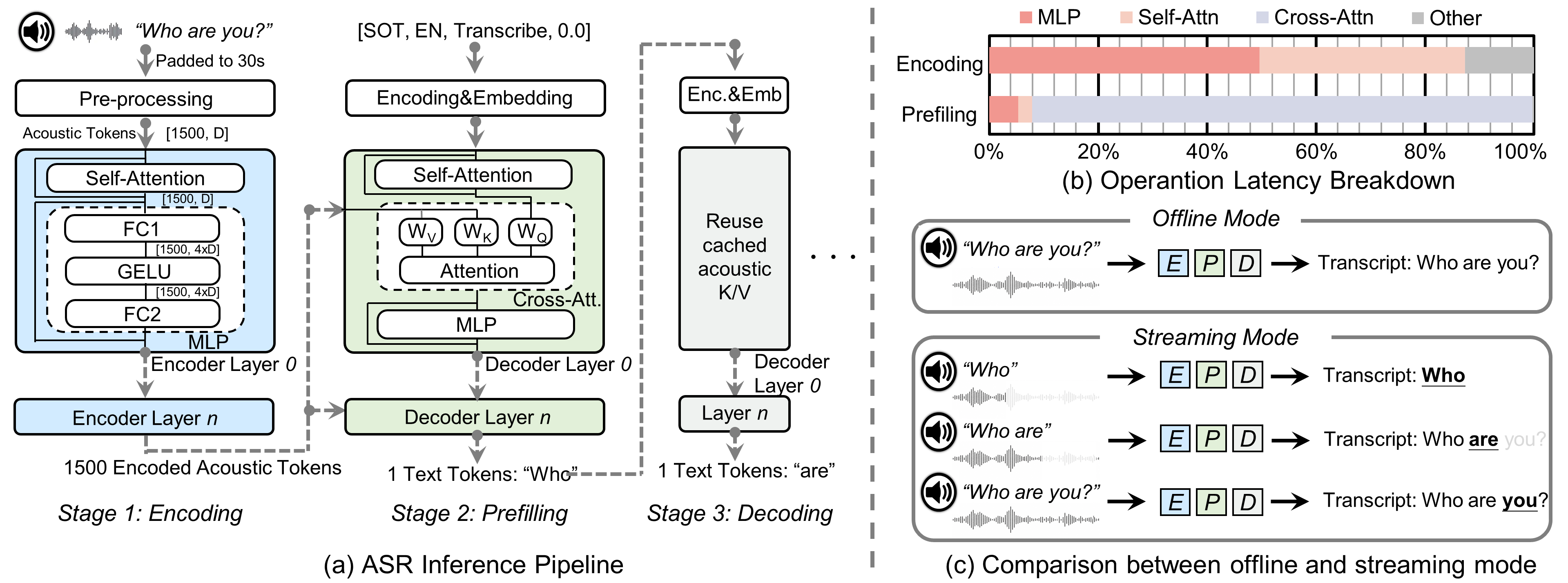}
    \caption{(a) Inference pipeline of a Whisper-style encoder--decoder ASR
    model, consisting of encoding (E), prefilling (P), and autoregressive decoding (D) stages.
    (b) Normalized latency breakdown for the encoding and prefilling stages
    in Whisper-Medium.
    (c) Comparison of offline and streaming modes. In streaming mode,
    encoding and prefilling are repeated as each new speech segment arrives.}
    \label{fig:asr-inference-workflow}
\end{figure*}

This cost stems from the long, dense sequences of acoustic tokens that ASR must
process. For example, Whisper adopts an encoder--decoder architecture in which
the encoder converts each fixed 30-second input window into 1,500 acoustic
tokens~\cite{radford2023robust}. As shown in Fig.~\ref{fig:asr-llm-token-imbalance}, for the same speech
context, the ASR token count is nearly 23 times the average LLM input-token count.
Such a large token gap results in substantially higher computation in ASR.
More importantly, this gap is further amplified in real-time transcription,
where speech arrives incrementally and must be processed repeatedly
~\cite{wang2025whisperflow,wang2024simul} (see
Section~\ref{sec:offline-streaming-asr}).
\IEEEpubidadjcol

Existing approaches reduce ASR inference cost in three ways.
\emph{First, model-centric optimization} retrains models to process short or
streaming utterances rather than fixed-length windows, avoiding computation on
padding~\cite{krichli2025carelesswhisper,wang2025whisperflow,puvvada2024less,jeffries2024moonshine}.
However, the computation for spoken content remains unchanged, and
retraining can limit the generality of pretrained models. \emph{Second, dynamic
token pruning} removes less informative tokens during inference~\cite{ren202618},
but pruning overhead can limit speedup~\cite{tuli2023acceltran}, and irregular
computation can reduce hardware utilization~\cite{kwon2023mobile}.
\emph{Third, per-token cost reduction}, such as
quantization~\cite{yang2022dtatrans,wang2024improving} and low-rank
compression~\cite{kamahori2025liteasr}, uniformly lowers the cost of each token.
However, achieving a good balance remains difficult: aggressive compression
hurts accuracy, whereas conservative compression brings little speedup (see
Section~\ref{sec:per-token-limits}). Taken alone, these approaches remain
insufficient for efficient ASR inference.

Token merging offers a promising alternative and has been widely used in vision
Transformers (ViTs)~\cite{bolya2022token,kim2024token,norouzi2024algm,bolya2023token}. Before compute-intensive operations, similar tokens are merged
into representative tokens. The subsequent dense operations then run only on
the representatives rather than the full token sequence. Unlike pruning, which
discards selected tokens, token merging combines similar tokens so that their
information is still retained~\cite{bolya2022token}. The original token layout can then be restored
when needed~\cite{norouzi2024algm,bolya2023token}. This is important for ASR because transcription relies heavily on
the temporal alignment of each acoustic token. Furthermore, token merging can be
applied at inference time without retraining or modifying model weights~\cite{kim2024token}, thereby
preserving the generality of pretrained multilingual ASR models.

Although token merging is effective, applying it directly to ASR faces two
challenges. \emph{Challenge (a): Uniform Merging Policy Across Layers.} Existing
methods either repeat token matching at every layer~\cite{yoo2026tri,shin2025atom} or reuse the matching
results from a previous layer~\cite{ren202618}. In encoder--decoder ASR, however, token
similarity changes across encoder layers but often remains stable across decoder
layers (see Section~\ref{sec:acoustic-token-similarity}). Repeating token
matching at every layer adds unnecessary matching overhead, whereas simply
reusing previous matching results can carry outdated token relations and hurt
transcription quality. \emph{Challenge (b): Tradeoff Between Matching Overhead
and Merging Ratio.} To reduce matching overhead, some methods partition tokens
into fixed windows and
compare tokens only within the same window~\cite{yoo2024adaptiv}. But in ASR, similar-token regions
vary with speech content. Fixed windows can miss tokens that
should be merged, lowering the merging ratio and limiting speedup. In contrast,
global token matching~\cite{bolya2022token} is more flexible but incurs prohibitive overhead.

To address these challenges, we propose \textit{MergeASR}, a
hardware--software co-designed token-merging framework for Whisper-style
encoder--decoder ASR. It improves the tradeoff between inference efficiency and
transcription quality across model sizes and transcription settings. Our main
contributions are as follows:

\smallskip
{\setlength{\leftskip}{\parindent}\noindent 1) We systematically analyze the within-layer distribution and cross-layer
persistence of token similarity in ASR and propose a component-aware merging
policy. MergeASR performs independent merging at each encoder layer, where
token similarity evolves from layer to layer, but reuses stable merging results
in the decoder, avoiding redundant matching and quality loss from outdated
token relations (see Sections~\ref{sec:acoustic-token-similarity} and
\ref{sec:merging-unmerging}).\par}

\smallskip
{\setlength{\leftskip}{\parindent}\noindent 2) We propose cache-based token matching, which exploits
the temporal locality of neighboring acoustic tokens. A small cache maintains a
few candidate representatives likely to match incoming tokens and dynamically
updates the set as speech content changes. It avoids restricting matching to
fixed windows or searching all tokens globally, while achieving a high merging
ratio (see Section~\ref{sec:temporal-local-matching}).\par}

\smallskip
{\setlength{\leftskip}{\parindent}\noindent 3) We design a dedicated, plug-and-play token-merging (TM) engine for
MergeASR. It integrates low-precision token matching, token-batch
pipeline scheduling, and compact cache management to reduce the additional
latency and hardware cost of token merging (see
Section~\ref{sec:mergeasr-hardware}).\par}

\smallskip
{\setlength{\leftskip}{\parindent}\noindent 4) To demonstrate the effectiveness of MergeASR, we build a cycle-level
hardware simulator and evaluate the framework across Whisper models of different
sizes. Experiments show that our design achieves up to 1.50$\times$
end-to-end speedup over the baseline with limited loss in transcription
quality. Energy profiling further shows 29.0\% energy
savings for the encoder part with our tailored token-merging design (see
Section~\ref{sec:evaluation}).\par}

\section{Background and Motivation}
\label{sec:background}

\subsection{ASR Inference Pipeline}
\label{sec:asr-inference-pipeline}

Modern ASR systems commonly adopt encoder--decoder
Transformers~\cite{puvvada2024less,jeffries2024moonshine}, such as
OpenAI's Whisper. Trained on large-scale, weakly supervised multilingual speech
data, Whisper delivers robust transcription across languages and acoustic
conditions~\cite{radford2023robust}.
As shown in Fig.~\ref{fig:asr-inference-workflow}(a), Whisper inference consists
of three stages. \emph{1) Encoding:} Whisper processes speech in fixed 30-s
windows, using feature extraction and downsampling to produce 1,500 temporally
ordered acoustic tokens. These tokens then pass through the encoder blocks to
form encoded acoustic tokens. \emph{2) Prefilling:} The decoder begins with a few task prompt
tokens specifying the language, task, and initial timestamp. During this
stage, each decoder block projects the
encoded acoustic tokens into cross-attention keys and values (K/V) and caches
them for later use. The initial transcription prediction is also produced at
this stage. \emph{3) Autoregressive decoding:} The decoder reuses the cached
acoustic K/V and predicts subsequent text tokens from the generated text prefix
until transcription is complete.

Fig.~\ref{fig:asr-inference-workflow}(b) shows the latency breakdown of the
compute-intensive encoding and prefilling stages during inference of
the Whisper-Medium model. The encoder multilayer perceptron (MLP) accounts for approximately
49.6\% of encoding latency, as it applies large linear
transformations to all 1,500 acoustic tokens. In prefilling, the bottleneck
shifts to cross-attention, which includes projecting the full encoded acoustic
sequence into K/V at each decoder layer and accounts for approximately 91.9\% of the
stage latency. The MLP and self-attention process only a few text tokens, with
a combined share below 10\%. Thus, although encoding and
prefilling serve different roles, both are dominated by dense computation over
acoustic tokens.

\vspace{-4pt}
\subsection{Offline and Streaming ASR}
\label{sec:offline-streaming-asr}

Encoder--decoder ASR typically relies on the full audio context for high-quality
transcription and therefore often processes complete utterances offline. This
offline setting, however, is difficult to adapt for most practical services,
which require streaming transcription as speech arrives. One straightforward
approach is to train a streaming-native model~\cite{ren202618} or fine-tune an
offline model for streaming~\cite{krichli2025carelesswhisper,wang2025whisperflow}.
However, supporting different languages, latency requirements, and
audio conditions may require repeated training or adaptation, complicating
deployment and increasing maintenance costs. Therefore, many studies try to
reuse existing offline models to enable streaming transcription without
additional training~\cite{wang2024simul,machavcek2023turning}.

As illustrated in Fig.~\ref{fig:asr-inference-workflow}(c),
this pseudo-streaming approach divides continuous speech into short audio
chunks. Each new chunk is appended to the accumulated audio, and ASR inference
is repeated on the full updated audio. The system outputs only new text and
uses previously generated text as a prompt for subsequent inference, avoiding
repeated output~\cite{wang2024simul}. This approach can preserve the offline model's transcription
quality and robustness, but also repeats the compute-intensive processing of
acoustic tokens. In this example, encoding and prefilling are each performed
three times for the same utterance in streaming mode, whereas offline
processing requires only one pass. Therefore, the substantial computational
cost of acoustic-token processing in both stages is further amplified.

\vspace{-4pt}
\subsection{Limits of Per-Token Optimization}
\label{sec:per-token-limits}

As discussed in Sections~\ref{sec:asr-inference-pipeline} and
\ref{sec:offline-streaming-asr}, acoustic tokens dominate the workload in the
compute-intensive stages of ASR inference. A natural approach is therefore to
reduce the processing cost of each token.
\emph{1) Activation quantization}~\cite{yang2022dtatrans} represents acoustic
activations with fewer bits, reducing computation and data movement.
\emph{2) Low-rank acoustic representation} maps acoustic tokens into a low-rank
subspace through offline analysis, uniformly reducing the feature dimension of
each token~\cite{kamahori2025liteasr}. In both cases, ASR models still need to
process the full acoustic-token sequence.

\begin{figure}[!t]
    \centering
    \includegraphics[width=\columnwidth]{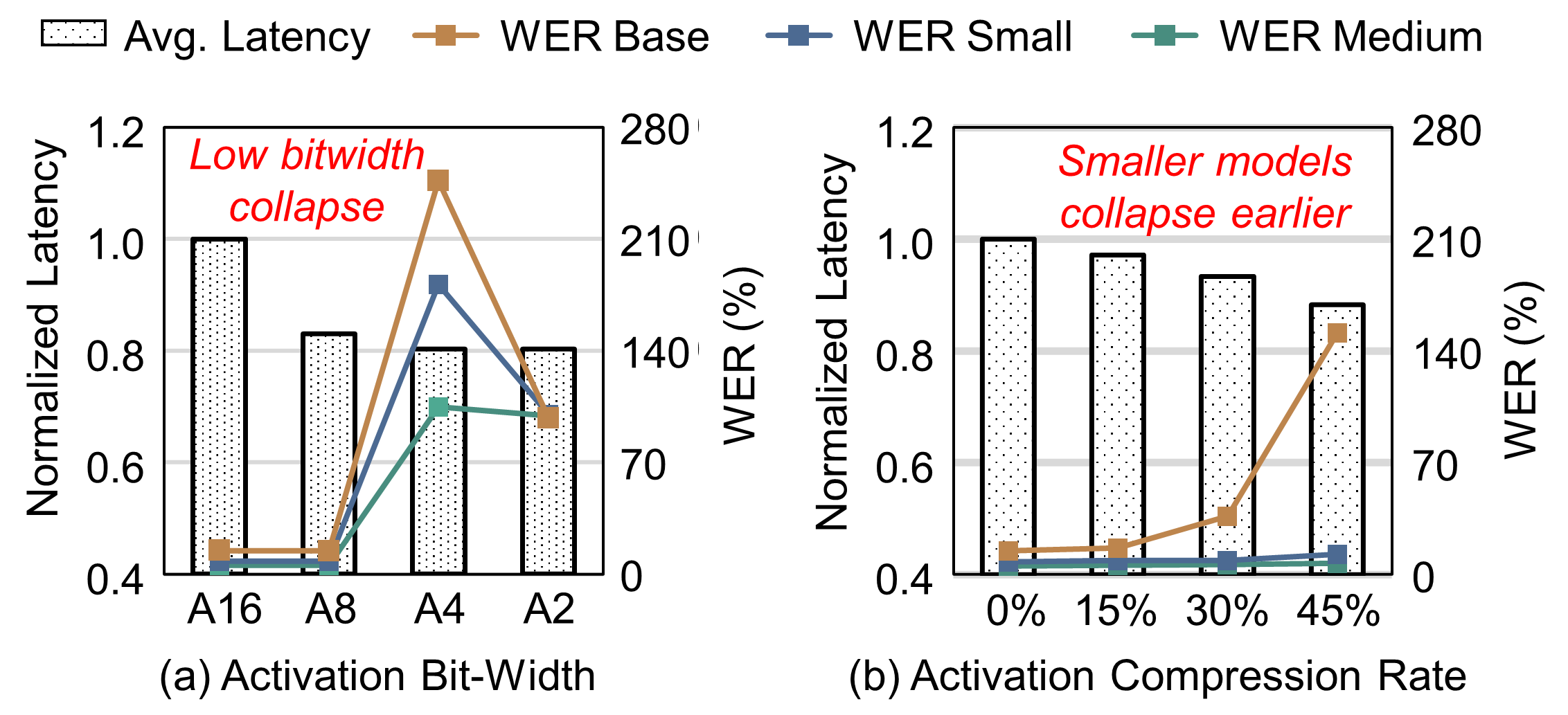}
    \caption{Accuracy--latency trade-offs of per-token optimization in
    Whisper-Base, Small, and Medium. (a) Activation quantization from 16 to
    2 bits. (b) Low-rank compression from 0\% to 45\%. Following
    LiteASR~\cite{kamahori2025liteasr}, both optimization and latency evaluation
    are limited to the encoder.}
    \label{fig:per-token-optimization-limits}
    \vspace{-12pt}
\end{figure}

Fig.~\ref{fig:per-token-optimization-limits} illustrates the limitations of
these approaches. Reducing activation precision from 16 to 8 bits leaves the
word error rate (WER) nearly unchanged. Further reduction to 4 or 2 bits
provides much smaller additional latency savings. However, WER rises sharply
and can exceed 100\%, making the transcription largely unusable. Low-rank
compression follows a similar pattern
(Fig.~\ref{fig:per-token-optimization-limits}(b)). More importantly, smaller
Whisper models begin to lose accuracy earlier, suggesting less tolerance for
dimension reduction. These results motivate a complementary direction:
reducing the number of tokens to be processed in compute-intensive ASR
operations.

\begin{figure*}[!t]
    \centering
    \includegraphics[width=\textwidth]{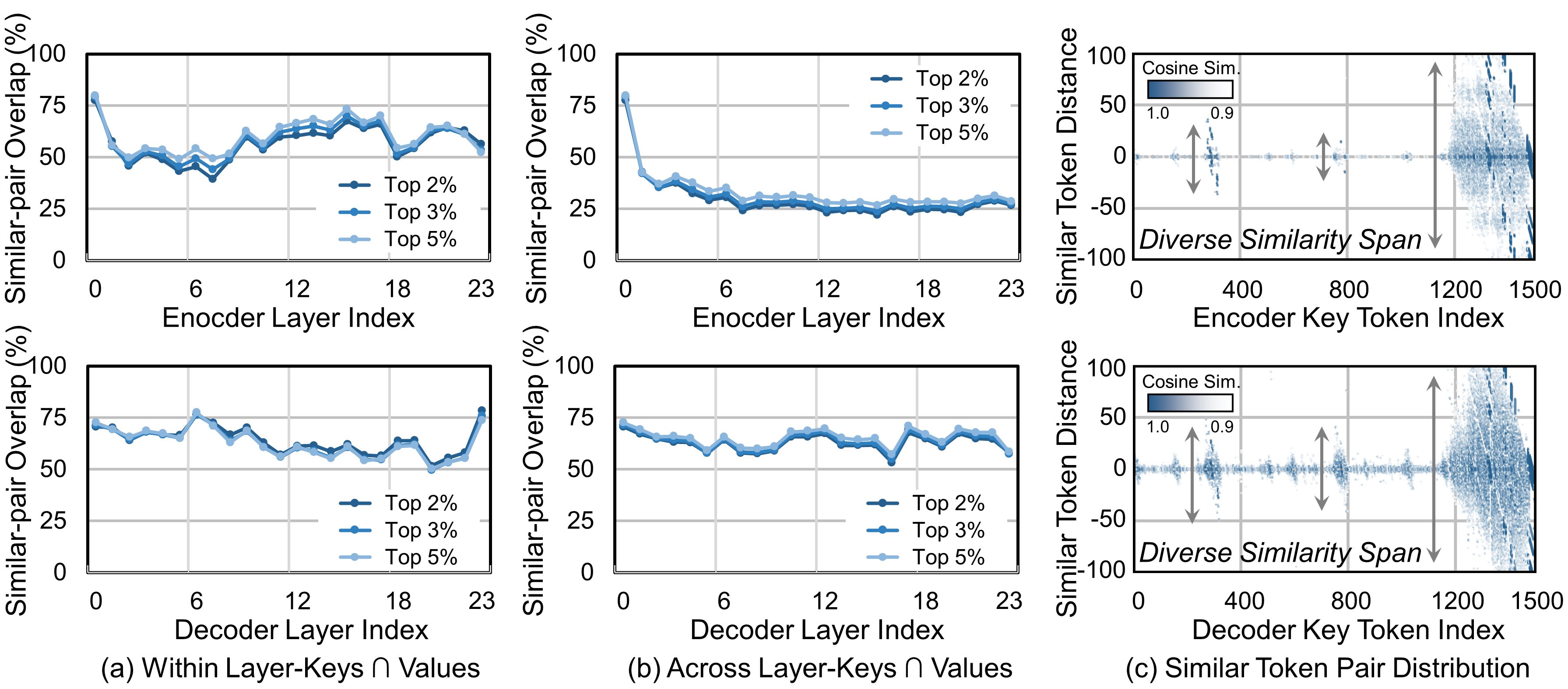}
    \caption{Acoustic-token similarity patterns. (a) Keys and Values show
    substantial overlap among their most similar token pairs within each layer.
    (b) Across layers, this overlap drops in the encoder as the layer index
    increases, but remains high in the decoder. Layer-0 Keys serve as the
    reference for Values across layers.
    (c) Similar tokens lie near the current token and form nearly continuous
    regions, but their span varies with speech content. Results are shown for
    layer 12, with similar token pairs defined by cosine similarity $> 0.9$.}
    \label{fig:acoustic-token-similarity}
    \vspace{-5pt}
\end{figure*}

\vspace{-4pt}
\subsection{Token Merging}
\label{sec:token-merging-background}

Token merging reduces repeated computation on tokens with similar
information. It first identifies similar tokens through \emph{token matching},
then replaces them with their weighted average or selects one of the similar
tokens as their representative~\cite{bolya2022token,kim2024token}. Subsequent operations
process the shorter sequence. In ToMe, merging occurs between attention and
the MLP in each ViT block, progressively reducing the
tokens processed by both operations across layers. This nearly doubles
inference throughput without retraining, with only a slight loss in
accuracy~\cite{bolya2022token}.

The computational savings from token merging, however, can be substantially
offset by token-matching overhead. To reduce this overhead, several
methods restrict matching candidates or simplify similarity
calculations~\cite{shin2025atom,yoo2024adaptiv}. Nevertheless, these
vision-oriented optimizations may not transfer directly to ASR. For example,
image tokens represent patches on a two-dimensional grid, whereas acoustic
tokens represent consecutive speech frames. Their similarity patterns and
locality can therefore differ. Moreover, ViTs typically use a homogeneous
encoder-only architecture, while ASR models use a heterogeneous
encoder--decoder architecture. A uniform merging policy may not suit all
these components. Therefore, exploiting the characteristics of acoustic
tokens for efficient matching remains a key challenge for token merging in
ASR.

\section{MergeASR Algorithm Optimization}
\label{sec:mergeasr-algorithm}

We first examine acoustic-token similarity to guide the design of MergeASR
(Section~\ref{sec:acoustic-token-similarity}). Based on these observations, we
describe how merging and unmerging are applied in the encoder and decoder
(Section~\ref{sec:merging-unmerging}). We then present cache-based matching
(Section~\ref{sec:temporal-local-matching}) and approximate center updates
(Section~\ref{sec:approximate-center-update}) to reduce matching overhead.

\vspace{-4pt}
\subsection{Observations on Acoustic-Token Similarity}
\label{sec:acoustic-token-similarity}

Token-merging studies on ViTs show that token similarity changes across layers.
A fixed merging ratio may therefore merge insufficiently similar tokens
in some layers while overlooking redundant tokens in others. These findings
motivate our analysis, but do not establish how token merging should be applied
to ASR. We examine acoustic-token similarity across tensors, across layers,
and along the token sequence. We use cosine similarity to measure how similar
two tokens are. We focus on Keys and Values because they summarize
the token information used by attention.

\subsubsection{Within-Layer Similarity}
Fig.~\ref{fig:acoustic-token-similarity}(a) examines whether similar token
pairs in Keys are also similar in Values within the same layer. We rank all
token pairs by cosine similarity separately in each tensor and retain the top
$p\%$ most similar pairs ($p\in\{2,3,5\}$). We then measure the overlap
between the selected pair sets. This overlap exceeds 50\% in many
encoder layers and reaches 75\% in some. The decoder shows similarly high
overlap. These results indicate that Keys and Values exhibit consistent
token-similarity patterns. This suggests that matching results can be reused
across tensors to guide token merging within the same layer.

\subsubsection{Cross-Layer Similarity}
\label{sec:cross-layer-similarity}
Fig.~\ref{fig:acoustic-token-similarity}(b) examines whether similar token
pairs persist across layers. We fix the top-$p\%$ pairs from layer-0 Keys as
the reference and compare them with the top-$p\%$ pairs from Values in each
layer. In the encoder, overlap drops rapidly as the layer index increases.
As hidden representations change across layers, tokens that were similar in an
early layer may no longer be similar later. Reusing early matching results can
therefore introduce incorrect merges. In contrast, overlap remains high in the
decoder. This is because its cross-attention Keys and Values in every layer
are derived from the same encoded acoustic sequence, which helps preserve
similarity across layers. This suggests that the encoder and decoder need
different policies for reusing matching results.

\subsubsection{Temporal Locality}
Fig.~\ref{fig:acoustic-token-similarity}(c) shows how far similar tokens lie
from the current token. We analyze encoder and decoder layer 12 using an
utterance shorter than 30 s, with the input padded to 30 s. For each acoustic
token, we plot the relative distances of other tokens with cosine similarity
above 0.9. Similar tokens tend to lie near the current token and form
nearly continuous regions, but the span of these regions varies with speech
content. Padding regions tend to have wider spans, while similarity in speech
regions is more localized and changes more quickly. Therefore, fixed
windows~\cite{norouzi2024algm} or partitions~\cite{bolya2022token} can waste
comparisons when the span is narrow and miss similar tokens when it is wide.
This suggests that acoustic token matching should adapt to the dynamic local
similarity span.

Overall, these observations motivate three design principles for MergeASR:
\begin{list}{\textbf{\theenumi}}{%
    \usecounter{enumi}%
    \renewcommand{\theenumi}{(\alph{enumi})}%
    \settowidth{\labelwidth}{\textbf{(c)}}%
    \setlength{\labelsep}{0.5em}%
    \setlength{\leftmargin}{\labelwidth}%
    \addtolength{\leftmargin}{\labelsep}%
    \addtolength{\leftmargin}{\parindent}%
    \setlength{\topsep}{3pt}%
    \setlength{\itemsep}{2pt}%
    \setlength{\parsep}{0pt}%
    \setlength{\partopsep}{0pt}%
}
    \item\label{design:within-layer-matching} Matching results from one
    tensor can guide token merging in other tensors within the same layer.
    \item\label{design:cross-layer-reuse} The encoder and decoder need
    different policies for reusing matching results across layers.
    \item\label{design:adaptive-matching} Token matching should adapt to
    the local similarity span as it changes with speech content.
\end{list}

\subsection{Merging and Unmerging}
\label{sec:merging-unmerging}

MergeASR adapts token merging and matching reuse to the distinct similarity
patterns of the ASR encoder and decoder.

\noindent\textbf{Within-Layer Merging and Unmerging.} For each token, we
average its Key vectors across attention heads to obtain a compact feature for
similarity matching~\cite{bolya2022token}. Tokens with similar features are grouped.
Following principle~\ref{design:within-layer-matching}, this grouping guides
merging across tensors within the same layer, avoiding repeated matching.

Given a token sequence $\boldsymbol{X}=[\boldsymbol{x}_1,\ldots,\boldsymbol{x}_T]$,
matching partitions the token indices into $M\leq T$ groups
$\{\mathcal{C}_m\}_{m=1}^{M}$. Each group is merged into one representative
token~\cite{bolya2022token}:
\begin{equation}
    \widetilde{\boldsymbol{x}}_m =
    \operatorname{Merge}\bigl(\{\boldsymbol{x}_i \mid i \in \mathcal{C}_m\}\bigr).
    \label{eq:merge}
\end{equation}
The representatives form the compressed sequence
$\widetilde{\boldsymbol{X}}=[\widetilde{\boldsymbol{x}}_1,\ldots,
\widetilde{\boldsymbol{x}}_M]$. The dense operation $f$, such as the encoder MLP,
processes these $M$ representatives instead of the original $T$ tokens,
producing $\widetilde{\boldsymbol{Y}}=f(\widetilde{\boldsymbol{X}})$. When the original
token layout is needed, unmerging copies each representative's output
$\widetilde{\boldsymbol{y}}_m$ back to the positions of the tokens in its group~\cite{bolya2023token}:
\begin{equation}
    \boldsymbol{y}_i = \widetilde{\boldsymbol{y}}_m,
    \qquad \forall i \in \mathcal{C}_m.
    \label{eq:unmerge}
\end{equation}
This restores the sequence length to $T$ while allowing tokens in the same
group to share the computed output.

\begin{figure}[!t]
    \centering
    \includegraphics[width=0.9\columnwidth]{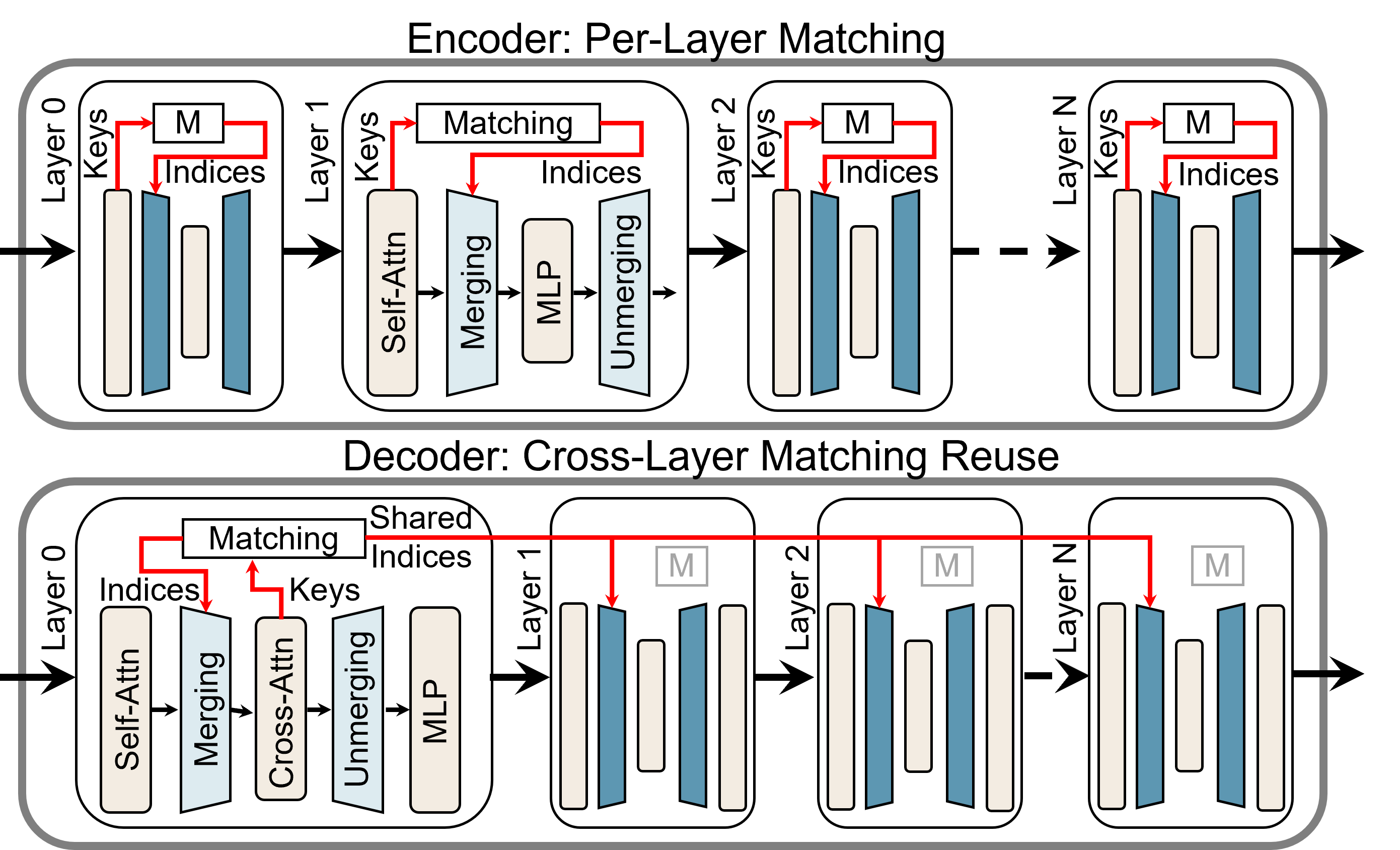}
    \caption{Component-specific token merging in MergeASR. The encoder
    performs matching independently at each layer to compress MLP inputs.
    The decoder reuses matching indices from the first layer to compress
    cross-attention Keys and Values across layers.}
    \label{fig:component-specific-merging}
\end{figure}

\noindent\textbf{Component-Specific Merging Policy.} MergeASR selects where
to merge tokens based on computation cost and when to repeat matching based on
cross-layer similarity. Following principle~\ref{design:cross-layer-reuse},
MergeASR uses different matching policies for the encoder and decoder, as shown
in Fig.~\ref{fig:component-specific-merging}. In the encoder,
merging compresses the input to each MLP, reducing a major source of encoding
computation. The output is then unmerged to restore the full token layout for
the next layer. Since token similarity changes across encoder layers,
MergeASR performs token matching
anew at each layer to preserve acoustic details that could be lost by reusing
earlier merging decisions. In the decoder, merging compresses the acoustic K/V
sequence used by cross-attention, the dominant operation during prefilling.
Since similarity patterns remain more stable across decoder layers, matching
is performed only in the first layer. Subsequent layers reuse this grouping
to merge their own Keys and Values, reducing cross-attention computation
without repeating token matching.

\begin{figure}[!t]
    \centering
    \includegraphics[width=\columnwidth]{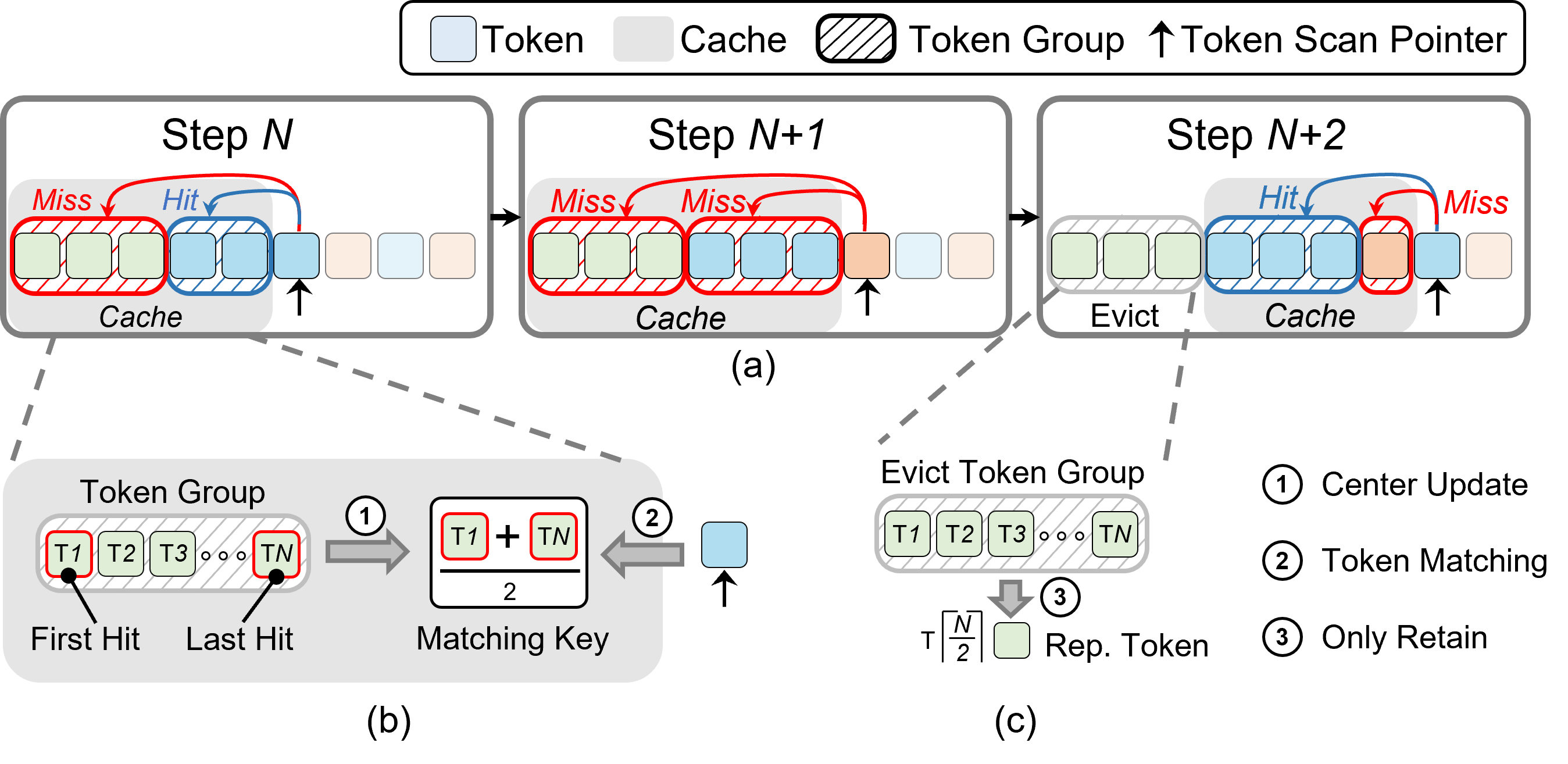}
    \caption{Cache-based token matching in MergeASR. (a) Each incoming token
    joins a matching group or starts a new group. (b) The group center is
    approximated by averaging the first and latest Keys. (c) The middle token
    in arrival order is selected as the representative when the group is
    evicted.}
    \label{fig:temporal-local-matching}
\end{figure}

\subsection{Cache-Based Token Matching}
\label{sec:temporal-local-matching}

Iterative clustering methods such as $k$-means~\cite{marin2023token} require
multiple rounds of token comparison and group updates, increasing matching
latency. ToMe~\cite{bolya2022token} instead uses
one-pass bipartite matching: it partitions $T$ tokens into source and
destination sets, compares every source token with every destination token,
and retains the most similar candidate pairs for merging. This avoids iterative
clustering but still requires $O(T^2)$ similarity comparisons. Fixed-window
matching~\cite{norouzi2024algm} further limits each token to nearby candidates and reduces the
complexity to $O(T)$. However, as indicated by
principle~\ref{design:adaptive-matching}, a fixed window cannot adapt
to the varying similarity span of acoustic tokens: a narrow window misses
mergeable tokens, whereas a wide one incurs redundant matching.

To avoid fixed window boundaries, MergeASR uses a small cache to retain
recently matched token groups. As shown in
Fig.~\ref{fig:temporal-local-matching}(a), it scans acoustic tokens once in
temporal order. Each cache entry stores a matching center for one group, and
each incoming token is compared with these centers. If a match is found, the
token joins that group and the entry becomes the most recently used.
Otherwise, the token starts a new group, replacing the least recently used
entry when the cache is full. A group can thus collect similar tokens
separated by intervening tokens, while groups that are no longer matched are
gradually replaced. This lets the candidate groups adapt to the speech
content. We use two cache entries, so each token requires at most two comparisons,
resulting in $O(T)$ matching complexity.

\begin{figure}[!t]
    \centering
    \begin{minipage}[t]{0.48\columnwidth}
        \centering
        \includegraphics[height=0.84\linewidth]{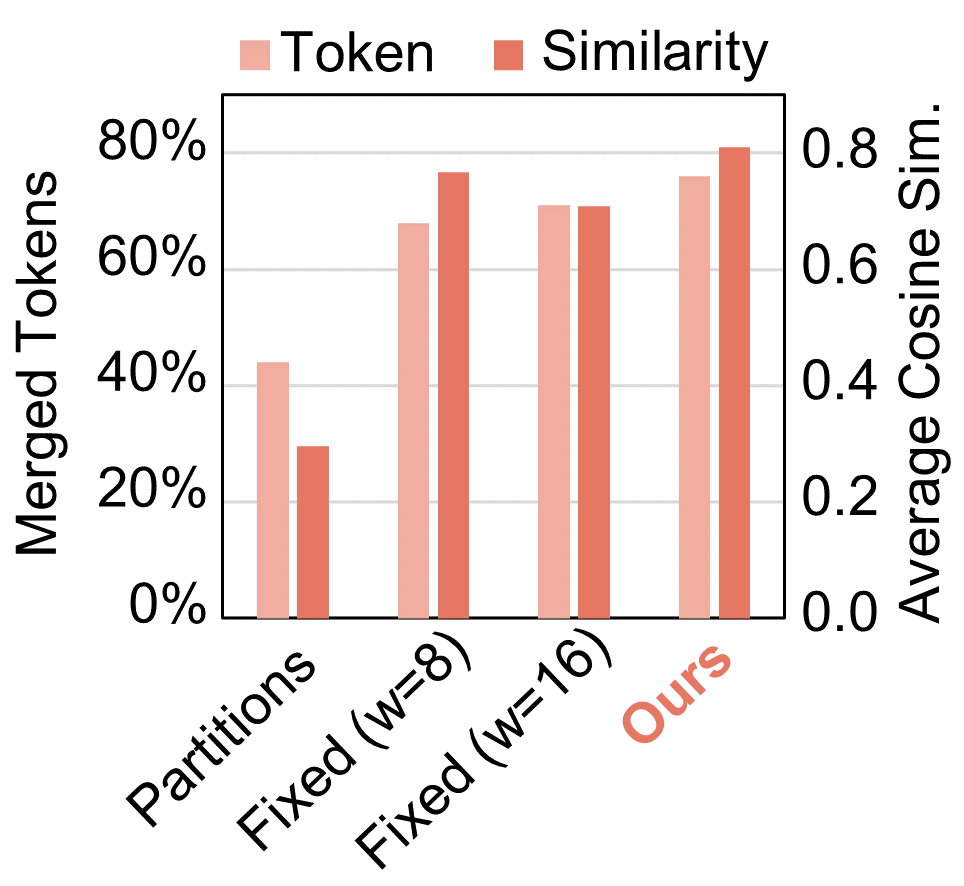}
        {\raggedright
        \hyphenpenalty=10000
        \captionof{figure}{\raggedright Comparison of token matching strategies.}
        \label{fig:matching-strategy-comparison}\par}
    \end{minipage}\hfill
    \begin{minipage}[t]{0.48\columnwidth}
        \centering
        \includegraphics[height=0.84\linewidth]{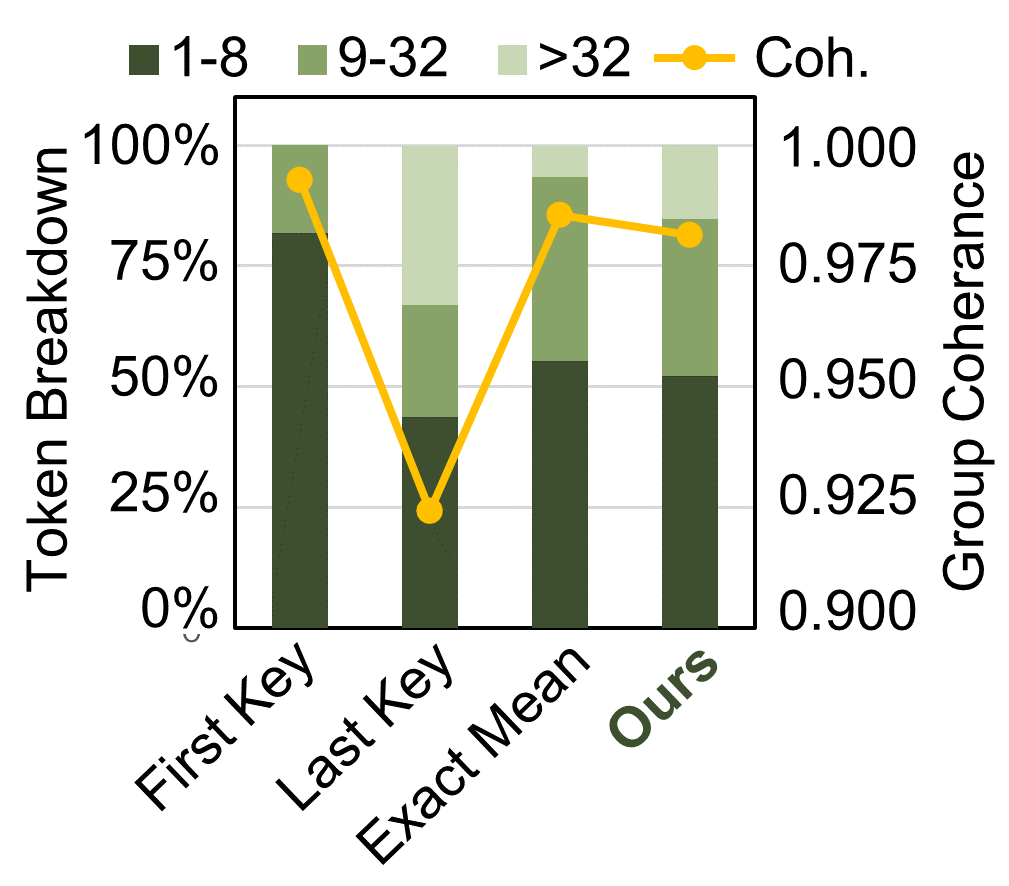}
        {\raggedright
        \hyphenpenalty=10000
        \captionof{figure}{\raggedright Comparison of matching center update methods.}
        \label{fig:matching-center-update}\par}
    \end{minipage}
\end{figure}

Fig.~\ref{fig:matching-strategy-comparison} compares the methods using two
metrics: the merged-token ratio measures token reduction, while the average
cosine similarity over all compared candidate pairs indicates how closely the
search focuses on similar tokens. Bipartite matching considers candidates
across the sequence, but its low average similarity indicates that comparisons
often involve less similar tokens. Fixed-window matching focuses on nearby
tokens. Enlarging the window size from $w=8$ to $w=16$ tokens enables more merging but also
introduces less similar candidates. Cache-based matching achieves both the
highest merged-token ratio and the highest average candidate similarity among
the evaluated settings. By retaining recently matched groups, it finds more
mergeable tokens while keeping comparisons focused on similar candidates.

\subsection{Approximate Center Update}
\label{sec:approximate-center-update}

Cache-based matching reduces comparisons, but leaves open how to create the
group centers. A natural way is to average all member Keys, as in
$k$-means~\cite{macqueen1967some}, and update the mean as tokens join. However,
maintaining a running sum requires a wider numerical range for larger groups.
Using each group's first token's Key
(\emph{First Key})~\cite{yoo2024adaptiv,shin2025atom} avoids these updates, but
the fixed center cannot reflect new members and may miss valid merges.
MergeASR therefore uses an approximate center update that accounts for new
members with low overhead.

MergeASR approximates each group center using the Keys of its first and most
recent members. For group $\mathcal{C}_m$, let these Keys be
$\boldsymbol{k}_m^{\mathrm{first}}$ and $\boldsymbol{k}_m^{\mathrm{last}}$.
As shown in Fig.~\ref{fig:temporal-local-matching}(b), the matching center is
\begin{equation}
    \widehat{\boldsymbol{c}}_m =
    \frac{\boldsymbol{k}_m^{\mathrm{first}} +
    \boldsymbol{k}_m^{\mathrm{last}}}{2}.
    \label{eq:approximate-center}
\end{equation}
When a new token joins, MergeASR updates the latest Key and recomputes this
average, keeping the first Key unchanged. The resulting center combines the
group's initial and latest information without accumulating all member Keys.
Since the averaged Key is used only for matching, MergeASR keeps the group's
middle token in arrival order as the representative $\widetilde{\boldsymbol{x}}_m$
when the group is evicted from the cache
(Fig.~\ref{fig:temporal-local-matching}(c)).

Fig.~\ref{fig:matching-center-update} compares four matching centers: the first
member's Key (First Key), the latest member's Key (Last Key), the mean of all
member Keys (Exact Mean), and the mean of the first and latest Keys (Ours).
The bars show how tokens are distributed across groups of different sizes,
while the line shows group coherence, which reflects average token similarity
within each group. With a fixed center, First Key leaves about 82\% of tokens
in small groups (1--8 tokens), indicating overly conservative merging.
In contrast, Last Key forms larger groups, but coherence drops to 0.92,
showing that the additional merging comes at the cost of merging less similar
tokens. Our method achieves a better balance: the latest Key allows the center
to change, while the first Key prevents it from simply following each new
member. This simple two-Key update yields a group-size distribution close to
Exact Mean and similarly high coherence of 0.98.

\section{MergeASR Hardware}
\label{sec:mergeasr-hardware}

\subsection{Architecture Overview}
\label{sec:hardware-architecture-overview}

\begin{figure}[!t]
    \centering
    \includegraphics[width=\columnwidth]{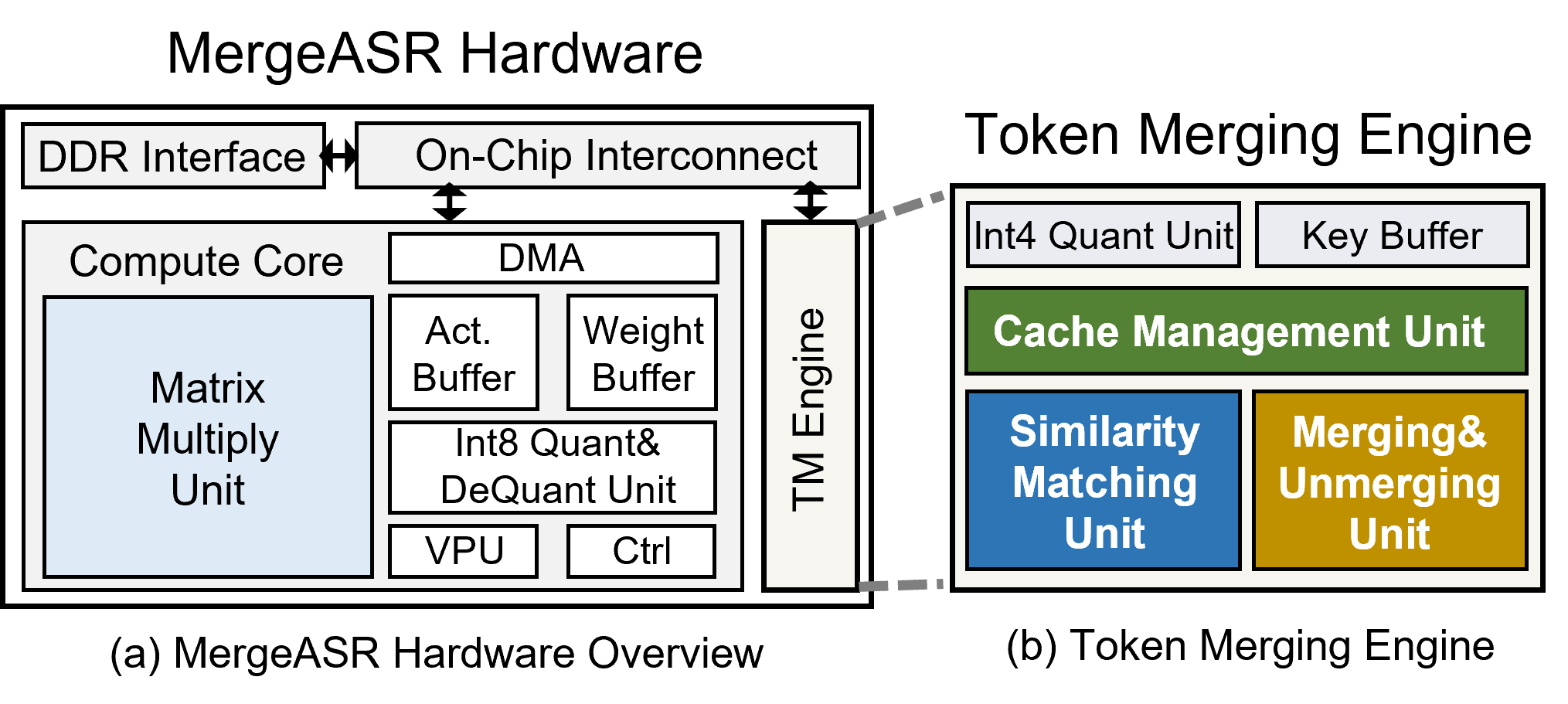}
    \caption{MergeASR hardware architecture: (a) overall architecture and
    (b) token-merging engine.}
    \label{fig:hardware-architecture}
\end{figure}

\noindent\textbf{Overview.} Figure~\ref{fig:hardware-architecture} shows
the MergeASR hardware architecture, consisting of a Compute Core, a
TM engine, an on-chip interconnect, and a DDR interface.
The Compute Core performs the main dense operations of the encoder--decoder
ASR model. The TM engine implements the proposed merging algorithm in
Section~\ref{sec:mergeasr-algorithm} to reduce the number of acoustic tokens
processed by these operations. Specifically, it receives attention Keys from
the Compute Core, identifies similar tokens, and produces a compressed token
sequence for subsequent computation. When the original token layout is needed,
it restores that layout through unmerging. The DDR interface supplies model
weights and intermediate data, while the on-chip interconnect carries data
between the components. Therefore, the Compute Core and TM engine can directly
exchange token data and group assignments on chip, reducing DDR traffic.

\smallskip
\noindent\textbf{Compute Core.} The Compute Core uses a matrix multiplication
unit for the main matrix operations in ASR, while a vector processing unit
(VPU) handles operations such as Softmax and GELU. Activation and weight
buffers supply data to these units. The DMA engine prefetches weights and
activations from DDR while computation proceeds, reducing delays caused by
data transfers. As shown in Section~\ref{sec:per-token-limits}, INT8 activation
quantization reduces computation cost while preserving transcription quality.
The Compute Core therefore includes an INT8 quantization and dequantization
unit to support low-precision computation. A control unit coordinates
computation and data transfers across these components.

\subsection{Token Merging Engine}
\label{sec:token-merging-engine}

\noindent\textbf{Overall Design.} The TM engine in
Fig.~\ref{fig:hardware-architecture}(b) is designed to reduce merging latency
with low area and on-chip memory overhead. Its INT4 Quant Unit first quantizes
head-averaged attention Keys and stores them in token order in the Key Buffer.
These quantized Keys guide token matching, while dense operations continue to
use INT8 activations. Section~\ref{sec:evaluation-results} shows that 4-bit Keys retain sufficient
similarity information for effective matching. This reduces matching hardware
cost and halves Key Buffer storage compared with INT8. After quantization,
the Similarity Matching Unit (SMU), Cache Management Unit (CMU), and Merging
and Unmerging Unit (MUU) perform group matching, cache updates, and token
merging and unmerging, respectively, as detailed below.

\begin{figure}[!t]
    \centering
    \includegraphics[width=\columnwidth]{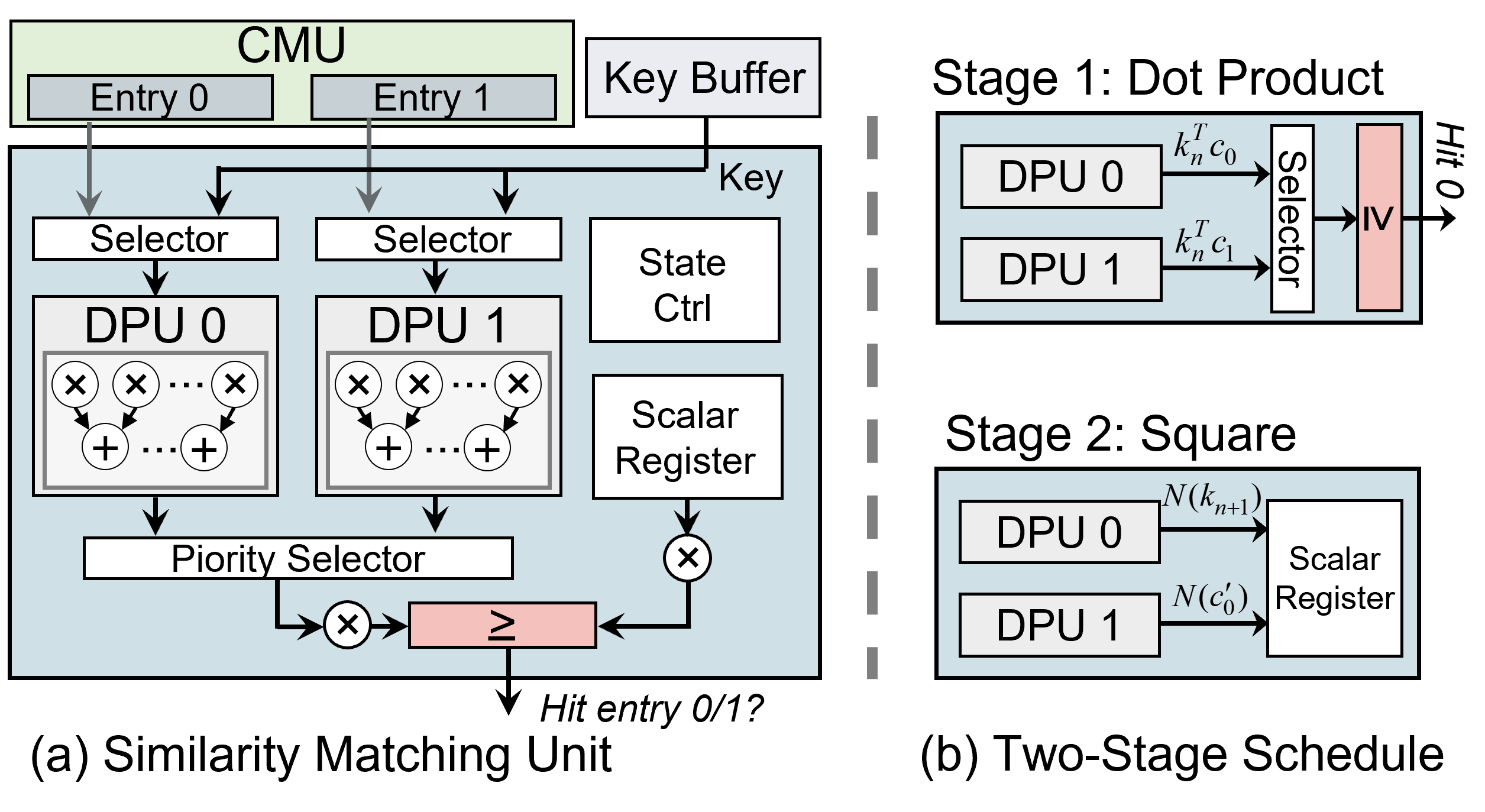}
    \caption{(a) Architecture overview of the Similarity Matching Unit (SMU).
    (b) Two-stage schedule for hit detection and parallel squared-norm
    computation. $N(\boldsymbol{x})=\lVert\boldsymbol{x}\rVert_2^2$ denotes
    the squared norm of $\boldsymbol{x}$.}
    \label{fig:similarity-matching-unit}
\end{figure}

\smallskip
\noindent\textbf{Similarity Matching Unit.} As shown in
Fig.~\ref{fig:similarity-matching-unit}(a), the SMU compares the current token's
Key $\boldsymbol{k}_n$ with the two cached group centers. A group with center
$\boldsymbol{c}_i$ is a hit when its cosine similarity with the current Key
is at least a threshold $\tau$:
\begin{equation}
    \frac{\boldsymbol{k}_n^{\mathsf{T}}\boldsymbol{c}_i}
    {\lVert\boldsymbol{k}_n\rVert_2\lVert\boldsymbol{c}_i\rVert_2}
    \geq \tau.
    \label{eq:cosine-similarity}
\end{equation}
To avoid costly division and square roots, the SMU reformulates this condition as
\begin{equation}
    \bigl(\boldsymbol{k}_n^{\mathsf{T}}\boldsymbol{c}_i\bigr)^2
    \geq \tau^2\lVert\boldsymbol{k}_n\rVert_2^2
    \lVert\boldsymbol{c}_i\rVert_2^2,
    \qquad \boldsymbol{k}_n^{\mathsf{T}}\boldsymbol{c}_i > 0.
    \label{eq:squared-similarity}
\end{equation}
Two dot-product units (DPUs) compute the dot products and squared norms required
by Eq.~\eqref{eq:squared-similarity}. The Scalar Register stores the squared
norms, which combine with $\tau^2$ to form the comparison threshold. The Priority
Selector selects a dot-product result according to cache priority. The selected
result is then squared and checked against the corresponding threshold by a
shared Threshold Comparator, avoiding a separate comparator for each group.

A key challenge in the SMU is the dependency between consecutive Keys. When
the current Key $\boldsymbol{k}_n$ hits an active group $i$, its matching center
changes from $\boldsymbol{c}_i$ to $\boldsymbol{c}'_i$. For the next Key
$\boldsymbol{k}_{n+1}$ to be matched, the updated center's squared norm,
$\lVert\boldsymbol{c}'_i\rVert_2^2$, must be ready. If the SMU updates the center
and computes its norm sequentially after every hit, the resulting delays
accumulate over a long acoustic-token sequence and reduce the performance gains
from token merging. To address this challenge, the SMU uses the two-stage
schedule in Fig.~\ref{fig:similarity-matching-unit}(b). In Stage~1,
the two DPUs compute $\boldsymbol{k}_n^{\mathsf{T}}\boldsymbol{c}_0$ and
$\boldsymbol{k}_n^{\mathsf{T}}\boldsymbol{c}_1$ for hit detection. Consider a hit
in Entry~0. In Stage~2, the CMU updates Entry~0's matching center using
Eq.~\eqref{eq:approximate-center} and streams $\boldsymbol{c}'_0$ to one DPU as
it is generated. This DPU computes $\lVert\boldsymbol{c}'_0\rVert_2^2$, while the
other computes the next Key's squared norm,
$\lVert\boldsymbol{k}_{n+1}\rVert_2^2$. Both norms are stored in the Scalar
Register for the next match. This schedule reduces the delay between
consecutive matches and improves DPU utilization.

\begin{figure}[!t]
    \centering
    \includegraphics[width=\columnwidth]{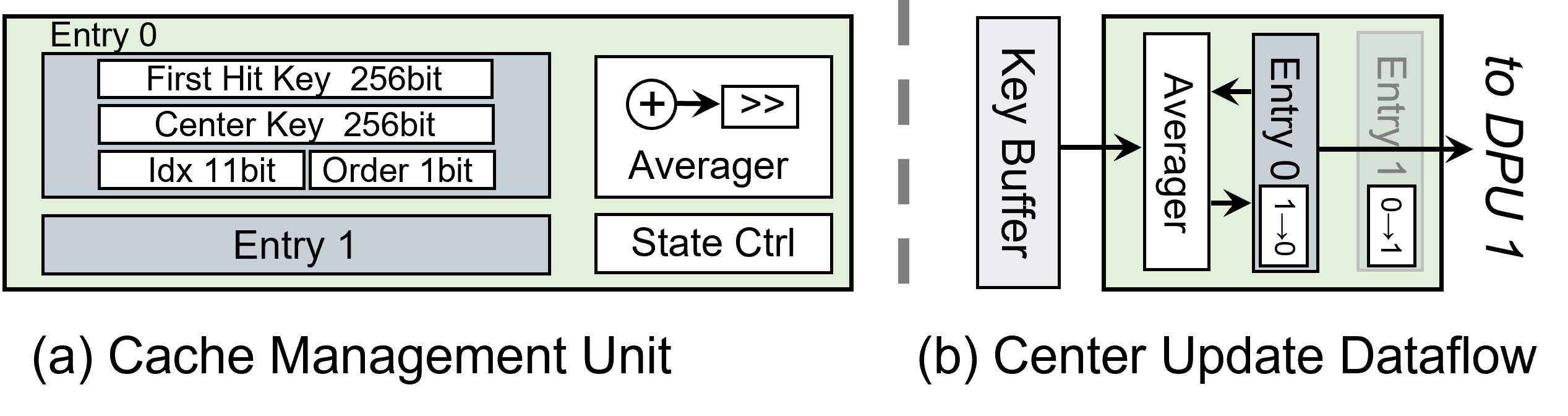}
    \caption{Cache Management Unit (CMU). (a) Architecture overview.
    (b) Dataflow for updating the matching center.}
    \label{fig:cache-management-unit}
\end{figure}

\smallskip
\noindent\textbf{Cache Management Unit.} After the SMU identifies a hit or
miss, the CMU updates the cache state used for matching. With the approximate
center update in Section~\ref{sec:approximate-center-update}, the CMU needs only
the stored First-Hit Key and the incoming Key to update a group's matching
center. As shown in Fig.~\ref{fig:cache-management-unit}(a), the CMU maintains
two active entries. Each entry stores a 256-bit First-Hit Key and a 256-bit
Center Key ($64\times4$ bits each), an 11-bit group index, and a 1-bit
cache-order flag. The index covers all groups in a 1,500-token Whisper sequence,
while the cache-order flag identifies which entry to evict when both are
occupied.

Fig.~\ref{fig:cache-management-unit}(b) shows how the CMU updates the matching
center on a hit. The Averager adds the stored First-Hit Key to the incoming Key
and shifts the sum right by 1 bit. Because all Keys are quantized using the
same symmetric scale, the Averager operates directly on their quantized values
without dequantization. Moreover, each update combines only two Keys, so only
1 extra bit per addition is required to prevent overflow, reducing the
hardware complexity. As the updated center is generated, it is written
directly to the selected entry and streamed to a DPU to compute the squared
norm needed in the two-stage schedule.

\begin{figure}[!t]
    \centering
    \includegraphics[width=\columnwidth]{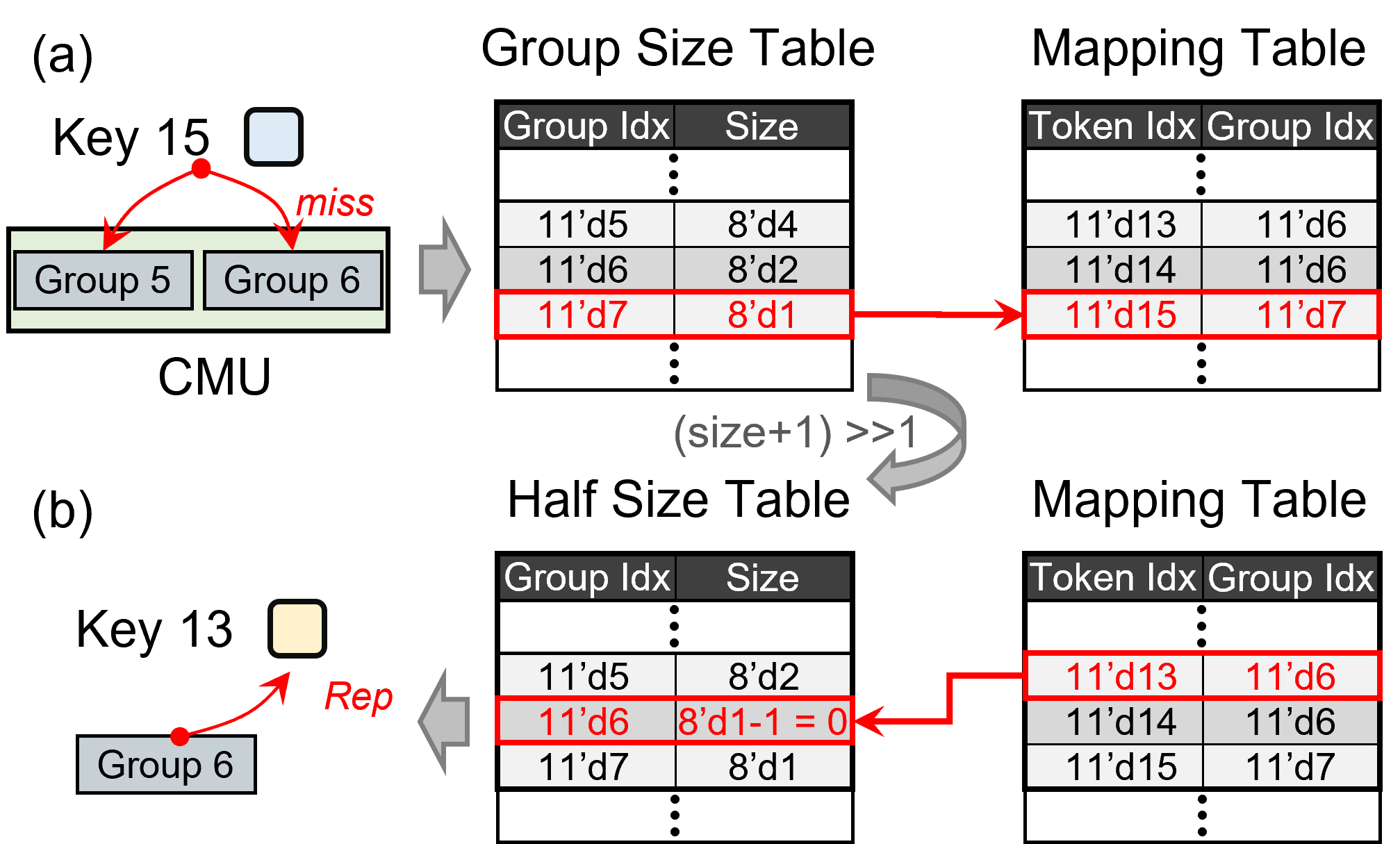}
    \caption{Group tables in the Merging and Unmerging Unit (MUU).
    (a) Recording token groups and their sizes.
    (b) Selecting the representative token of each group.}
    \label{fig:compact-metadata-management}
\end{figure}

\smallskip
\noindent\textbf{Merging and Unmerging Unit.} The CMU assigns each token to a
group, but simply storing every member index in its two cache entries would make each
entry as large as the largest possible group. The MUU instead records this
information in two fixed-size on-chip tables. The Mapping Table stores each
token's group index, and the Group Size Table stores the number of tokens for
each group. This keeps the cache entries small while preserving the information
needed for merging and unmerging operation.

\noindent\hspace*{\parindent}\textbf{(1) Recording Token Groups.} During token
matching, once the CMU determines an incoming token's group index, the MUU
writes its group index to the Mapping Table and increments the corresponding
count in the Group Size Table, as shown in
Fig.~\ref{fig:compact-metadata-management}(a). To further reduce the storage
overhead, we limit the maximum size of each group to 255 tokens, so only 8
bits are needed to store its size.

\noindent\hspace*{\parindent}\textbf{(2) Merging and Unmerging.} Once all
tokens have been assigned, the MUU converts each group size $s_m$ to
$\lceil s_m/2\rceil$. It then scans the Mapping Table in token order and counts
down a group's counter each time one of its tokens appears. The token that
brings the count to zero becomes that group's representative, as shown in
Fig.~\ref{fig:compact-metadata-management}(b). This scan completes merging.
For unmerging, the MUU reads the same table after the dense operation and
copies the outputs back.

\subsection{Streaming Token Matching}
\label{sec:token-level-pipeline-scheduling}

MergeASR's iterative matching assigns each token as soon as its head-averaged
Key is available. Each Key is compared with only a few active groups, so the TM
engine need not wait for the full Key sequence. The Compute Core generates Keys
in token order and sends them directly to the TM engine while continuing to
compute later tokens. Matching thus overlaps with Key generation, hiding much
of its latency and avoiding an extra off-chip write and read.

Representative selection, however, must wait until all group assignments are
known. In the encoder, it overlaps with the remaining attention computation
and finishes before the MLP begins. In the decoder, it delays the first
cross-attention, but subsequent layers reuse the first layer's group
assignments and representative indices, so they do not repeat matching or
selection.

\section{Evaluation}
\label{sec:evaluation}

\begin{figure*}[!t]
    \centering
    \includegraphics[width=\textwidth]{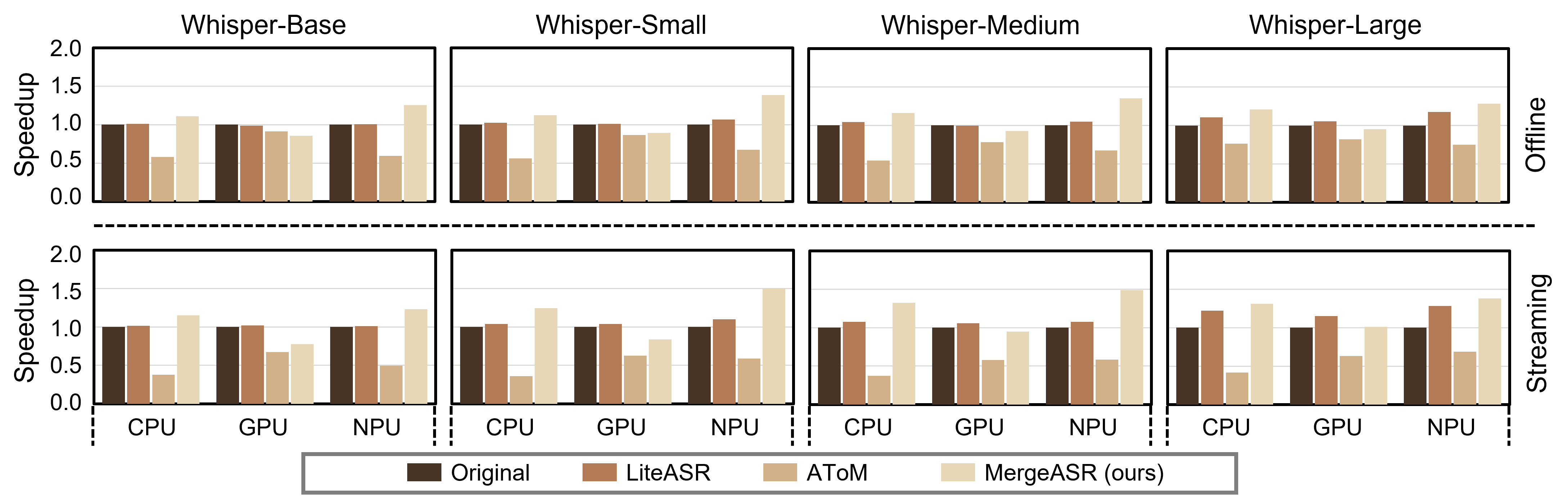}
    \caption{Speedup across the Whisper family on CPU, GPU, and NPU platforms
    in offline and streaming modes, normalized to the corresponding original
    model. We use a batch size of one for both modes and set the audio chunk
    size to 1~s in streaming mode. For NPU evaluation, linear operations use
    8-bit weights quantized with GPTQ~\cite{frantar2023gptq} and 8-bit
    activations (W8A8).}
    \label{fig:end-to-end-speedup}
\end{figure*}

\begin{table*}[!t]
\centering
\caption{Transcription accuracy and MAC reduction across Whisper models in offline and streaming modes.}
\label{tab:baseline-accuracy-mac}
\label{tab:accuracy-encoder-merging}
\scriptsize
\setlength{\tabcolsep}{1.75pt}
\renewcommand{\arraystretch}{1.18}
\begin{tabularx}{\textwidth}{@{}>{\centering\arraybackslash}p{0.52in}|>{\centering\arraybackslash}p{0.54in}|*{4}{D}|*{4}{D}|*{4}{D}|*{3}{D}@{}}
\hline
\multirow{3}{*}{\textbf{Setting}} & \textbf{Dataset}
& \multicolumn{4}{c|}{\textbf{LibriSpeech (test-clean)}}
& \multicolumn{4}{c|}{\textbf{LibriSpeech (test-other)}}
& \multicolumn{4}{c|}{\textbf{LibriSpeech}}
& \multicolumn{3}{c}{\textbf{MLS (test set)}} \\
\cline{2-17}
& \textbf{Metric}
& \multicolumn{4}{c|}{Accuracy (\%)\(\uparrow\)}
& \multicolumn{4}{c|}{Accuracy (\%)\(\uparrow\)}
& \multicolumn{4}{c|}{MAC Reduction (\%)\(\uparrow\)}
& \multicolumn{3}{c}{Accuracy (\%)\(\uparrow\)} \\
\cline{2-17}
& \textbf{Model} & B & S & M & L & B & S & M & L & B & S & M & L & B & S & M \\
\hline
\multirow{6}{*}{\textbf{Offline}} & \textbf{Original} & \textbf{94.83} & \textbf{96.52} & \textbf{97.12} & \textbf{97.26} & \textbf{87.73} & \textbf{92.44} & \textbf{94.09} & \textbf{94.97} & \textbf{0.00} & \textbf{0.00} & \textbf{0.00} & \textbf{0.00} & \textbf{71.96} & \textbf{83.69} & \textbf{88.60} \\
 & AToM & 85.27 & 95.01 & 96.68 & 95.10 & 74.14 & 88.41 & 93.61 & 91.64 & 35.18 & 35.98 & 24.75 & 33.61 & 66.98 & 82.26 & 87.64 \\
 & AdapTiV & 93.74 & 96.50 & 97.13 & 97.27 & 84.47 & 91.70 & 93.87 & 94.85 & 35.62 & 35.48 & 35.59 & 28.41 & 67.12 & 82.59 & 88.16 \\
 & ToMe & 93.96 & 96.13 & 96.77 & 96.55 & 85.67 & 90.92 & 92.76 & 93.24 & 18.50 & 21.28 & 22.79 & 23.79 & 67.56 & 80.69 & 86.56 \\
 & LiteASR & 93.39 & 96.21 & 97.11 & 94.83 & 82.29 & 90.69 & 93.61 & 91.90 & 11.05 & 17.53 & 20.30 & 35.75 & 29.71 & 73.17 & 87.02 \\
 & \textbf{Ours} & \textbf{84.71} & \textbf{95.49} & \textbf{95.26} & \textbf{97.05} & \textbf{77.76} & \textbf{90.56} & \textbf{91.77} & \textbf{94.47} & \textbf{34.29} & \textbf{37.99} & \textbf{40.97} & \textbf{36.21} & \textbf{66.09} & \textbf{79.34} & \textbf{87.28} \\
\hline
\multirow{6}{*}{\textbf{Streaming}} & \textbf{Original} & \textbf{92.48} & \textbf{95.66} & \textbf{97.10} & \textbf{96.76} & \textbf{82.27} & \textbf{89.54} & \textbf{93.05} & \textbf{91.77} & \textbf{0.00} & \textbf{0.00} & \textbf{0.00} & \textbf{0.00} & \textbf{64.43} & \textbf{79.57} & \textbf{86.19} \\
 & AToM & 84.61 & 91.67 & 96.90 & 92.86 & 75.07 & 83.12 & 92.16 & 83.04 & 33.55 & 36.17 & 23.15 & 37.82 & 57.39 & 71.56 & 85.16 \\
 & AdapTiV & 89.24 & 95.06 & 96.94 & 96.80 & 74.37 & 87.42 & 92.33 & 91.93 & 37.85 & 37.70 & 36.97 & 31.73 & 57.16 & 77.46 & 85.52 \\
 & ToMe & 90.70 & 94.83 & 96.59 & 95.60 & 78.80 & 86.52 & 91.27 & 88.64 & 18.63 & 21.40 & 22.98 & 23.92 & 58.37 & 75.23 & 83.47 \\
 & LiteASR & 89.75 & 94.65 & 96.97 & 94.87 & 72.94 & 85.06 & 92.13 & 87.29 & 11.12 & 17.62 & 20.41 & 35.97 & 19.35 & 65.05 & 84.24 \\
 & \textbf{Ours} & \textbf{90.96} & \textbf{95.34} & \textbf{96.96} & \textbf{96.69} & \textbf{78.72} & \textbf{88.39} & \textbf{92.67} & \textbf{91.75} & \textbf{34.55} & \textbf{38.86} & \textbf{41.75} & \textbf{39.06} & \textbf{60.44} & \textbf{77.16} & \textbf{85.61} \\
\hline
\end{tabularx}
\end{table*}

\subsection{Methodology}
\label{sec:evaluation-methodology}

\noindent\textbf{Models and Datasets.} We evaluate four multilingual Whisper
models~\cite{radford2023robust}: Base (74M parameters), Small (244M), Medium
(769M), and Large-v2 (1.55B, denoted Large). Tiny (39M) is omitted because of its limited
transcription quality~\cite{wang2025whisperflow}. For English evaluation, we
use the test-clean and test-other splits of LibriSpeech, an English read-speech
corpus~\cite{panayotov2015librispeech}, covering different levels of recognition
difficulty. We also use the Italian, Portuguese, and Polish test sets of
Multilingual LibriSpeech (MLS)~\cite{pratap2020mls} to assess generalization
beyond English. We measure transcription quality using WER,
a standard ASR metric that normalizes word substitutions, deletions, and
insertions by the number of reference words~\cite{radford2023robust,pratap2020mls}.
For clarity, transcription accuracy is reported as
$(1-\mathrm{WER})\times100\%$, where higher values indicate better performance.

\smallskip
\noindent\textbf{Inference Modes.} To cover different inference scenarios, we
evaluate MergeASR in offline and streaming modes, as illustrated
in Fig.~\ref{fig:asr-inference-workflow}(c). Offline inference transcribes a
complete utterance using the standard Whisper decoding
framework~\cite{radford2023robust}.
Streaming inference processes each
utterance in 1-s chunks using Simul-Whisper~\cite{wang2024simul}, which combines
cross-attention-guided decoding with truncation detection at chunk boundaries
to approach offline transcription quality without fine-tuning the pretrained
model. Both modes use batch size one and greedy decoding. To limit excessive
repetition in transcriptions, we use the same heuristic as
in~\cite{jeffries2024moonshine} and limit the output to at most six tokens per
second of audio. For the evaluation of speedup, we measure the complete
transcription time per utterance in offline mode and sum the processing times
of all its audio chunks in streaming mode.

\smallskip
\noindent\textbf{Hardware Platforms.} To evaluate end-to-end ASR performance,
we develop a cycle-level simulator of the MergeASR accelerator shown in
Fig.~\ref{fig:hardware-architecture}. The accelerator is configured at
1~GHz with 8.2~TOPS of INT8 throughput, 1-MB activation and weight buffers
each, and 51.2~GB/s of LPDDR4 bandwidth. The VPU is configured with 32 lanes,
and the DPUs in the TM engine with four lanes each. For comparison,
we define a baseline neural processing unit (NPU) by removing the TM
engine from MergeASR. For both configurations, we use a compiler to map ASR
workloads onto the hardware. To improve data reuse, the compiler automatically
tiles computation and manages on-chip data residency to reduce DRAM traffic.

To evaluate the algorithm alone, we also use an Intel Xeon Gold 6430 CPU with
32 physical cores and an NVIDIA GeForce RTX 4090 D GPU with approximately
294~TFLOPS of peak throughput and 1008~GB/s of GDDR6X
bandwidth. All CPU/GPU
implementations use PyTorch scaled dot-product attention (SDPA) for encoder
attention to provide optimized baselines for speedup comparisons. Decoder
attention remains explicit to support streaming alignment. For stable timing,
we warm up each implementation and report median latency over repeated runs.

\smallskip
\noindent\textbf{Baselines.} For accuracy evaluation, we compare MergeASR with
Original Whisper, LiteASR~\cite{kamahori2025liteasr}, AToM~\cite{shin2025atom},
AdapTiV~\cite{yoo2024adaptiv}, and ToMe~\cite{bolya2022token}. LiteASR applies
low-rank compression to token features, and we use its efficiency-focused
configuration. The other three methods were originally developed for
ViTs and typically pass merged tokens to subsequent layers. As
discussed in Section~\ref{sec:acoustic-token-similarity}, token similarity varies
across ASR encoder layers. For ASR, we therefore restore the full token layout
and match tokens independently in each encoder layer. For speedup evaluation,
we compare against Original Whisper, LiteASR, and AToM on CPU, GPU, and NPU
platforms. LiteASR's projection and reconstruction costs are both included.
Notably, AToM uses a 32-token matching window, and we map its matching
operations to the 32-lane VPU on the baseline NPU.

\smallskip
\noindent\textbf{Thresholds.} MergeASR uses model-specific, layer-wise encoder
thresholds and a fixed threshold of 0.927 across decoder layers. ToMe controls
token reduction through a merging ratio, which we set to 50\% for encoder MLP
inputs. AToM's encoder-attention, MLP, and decoder-cross-attention thresholds
are set to 1.5, 0.925, and 1.0, respectively. AdapTiV's encoder sign-similarity
threshold is set to 0.90. These settings remain fixed throughout the main
comparisons. To examine the impact of parameter choices, we also provide
results under varying merging parameters, as shown in
Fig.~\ref{fig:accuracy-merging-tradeoff}.

\begingroup
\clubpenalty=10000
\smallskip
\noindent\textbf{Area and Energy.} We synthesize the accelerator using
Design Compiler in TSMC 28nm, with compiler-generated SRAM macros.
On-chip energy is estimated using Power Compiler with VCS-generated
switching activity and simulated execution statistics. DDR energy is estimated
following~\cite{song20151}, using memory-access traces generated by our
simulator.
\par
\endgroup

\subsection{Experimental Results}
\label{sec:evaluation-results}

\noindent\textbf{Speedup.} Fig.~\ref{fig:end-to-end-speedup} compares
inference speedups across Whisper models and hardware platforms. On the NPU,
MergeASR achieves average speedups of 1.32$\times$ offline and 1.40$\times$ in
streaming mode, reaching 1.50$\times$ for Small. It consistently outperforms
the other baselines across the evaluated NPU configurations. These gains come
from lightweight matching and dedicated hardware support. LiteASR offers
smaller gains due to the additional cost of low-rank projection. AToM incurs
high matching overhead within its large fixed windows, offsetting the benefits
from token reduction.

MergeASR also accelerates CPU inference, achieving an average speedup of
1.26$\times$ in streaming mode. Both CPU and NPU generally benefit more in
streaming mode, where acoustic tokens are repeatedly processed during encoding
and prefilling. On the GPU, however, iterative matching introduces inter-token
dependencies that limit parallel execution, causing most configurations to
remain slower than Original. The MergeASR hardware maps this sequential
matching workload to a dedicated engine, thereby maintaining high utilization
of the compute core.

\smallskip
\noindent\textbf{Accuracy and MAC Reduction.}
\label{sec:accuracy-encoder-merging}
Table~\ref{tab:baseline-accuracy-mac} reports transcription accuracy and the
reduction in multiply--accumulate (MAC) operations across Whisper models and
inference modes. MAC reduction provides a common metric
for token merging and LiteASR's low-rank compression, with additional matching
MACs included. MergeASR reduces MACs by an average of 37.36\% offline and
38.56\% in streaming mode, reaching 41.75\% for Medium. It achieves the largest
reductions for Small, Medium, and Large in both modes, as cache-based matching
finds more mergeable tokens with low overhead.

In streaming mode, average accuracy losses relative to Original are 0.51 and
1.27 percentage points on test-clean and test-other, respectively. MergeASR
achieves both the highest accuracy among compressed methods and the largest
MAC reduction for Small and Medium on streaming test-other. Offline inference
uses the same acoustic features for the entire transcript, so merging errors
may persist throughout decoding and contribute to the larger accuracy loss.
In contrast, streaming refreshes these features and token groups with each
new audio chunk. Stricter merging thresholds in offline mode can help reduce
this loss. On MLS, MergeASR also achieves the highest streaming accuracy among
compressed methods for Base and Medium. These results suggest that online
token merging helps preserve the ASR model's cross-lingual generalization
without additional language-specific tuning.

\begin{figure}[!t]
    \centering
    \includegraphics[width=\columnwidth,trim=0 0 0 8.5bp,clip]{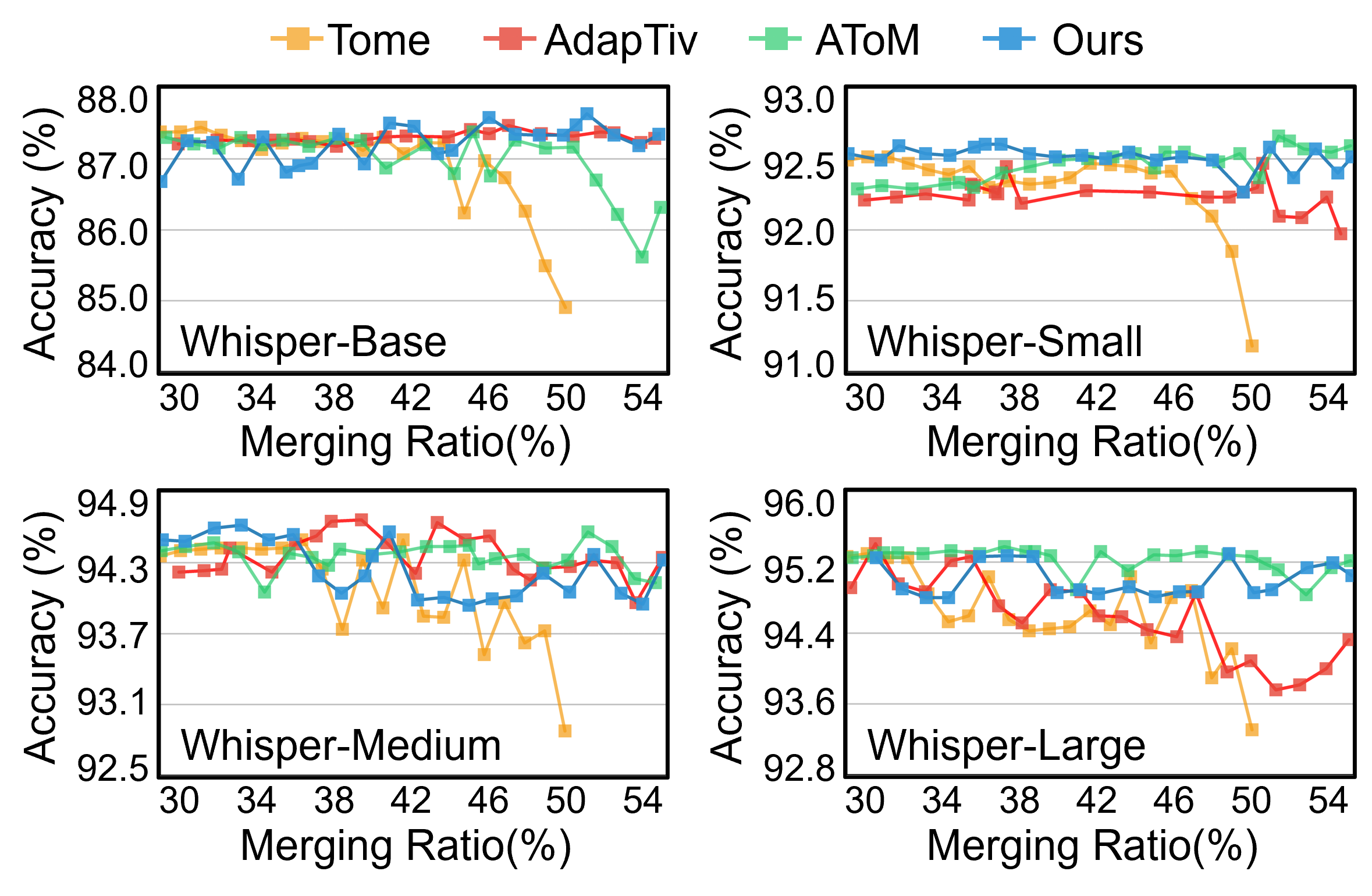}
    \caption{Trade-off between transcription accuracy and token merging ratio on LibriSpeech
    test-other. For a fair comparison, all methods apply token merging only to
    encoder MLPs and use offline inference.}
    \label{fig:accuracy-merging-tradeoff}
\end{figure}

\smallskip
\noindent\textbf{Accuracy--Merging Ratio Trade-off.}
Fig.~\ref{fig:accuracy-merging-tradeoff} shows how transcription accuracy
changes with the merging ratio. For a controlled comparison, all methods
apply merging only to encoder MLPs. We use offline inference on the more
challenging LibriSpeech test-other split. The merging ratio is averaged across
encoder layers. We sweep the thresholds for MergeASR,
AToM, and AdapTiV, while specifying ToMe's merging ratio directly. MergeASR
maintains relatively stable accuracy even at merging ratios above 50\%,
whereas ToMe degrades sharply near 50\%. AdapTiV loses accuracy at high ratios
for Large. AToM remains competitive for Small and Large but shows noticeable
degradation for Base at high ratios. The results show that MergeASR provides
a better balance between accuracy and token merging ratio, enabling high-quality,
computationally efficient transcription.

\begin{figure}[!t]
    \centering
    \includegraphics[width=\columnwidth]{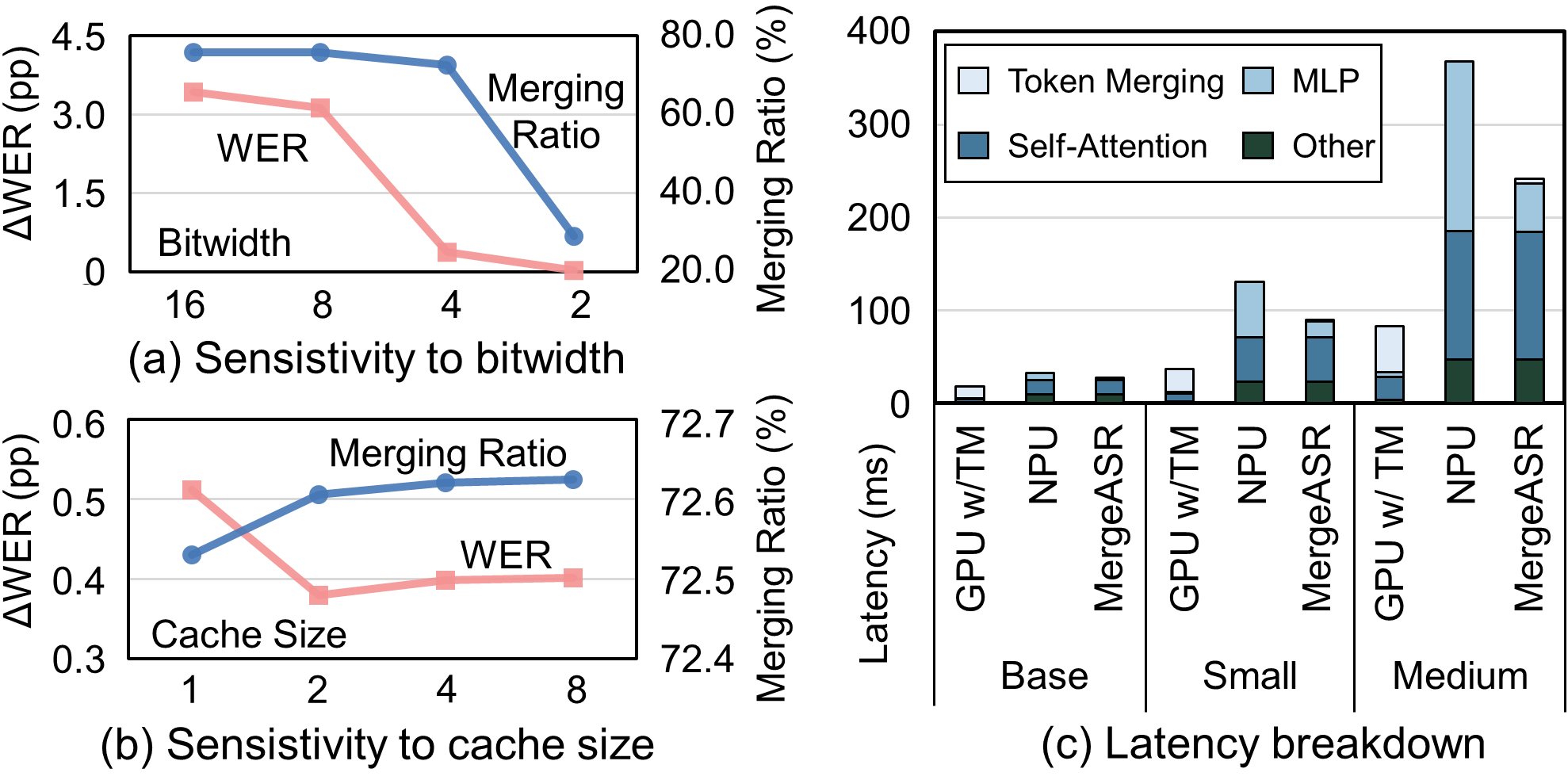}
    \caption{$\Delta$WER and token merging ratio with varying (a) Key bitwidth
    and (b) cache size. (c) Encoder latency breakdown for GPU with token
    merging, NPU, and MergeASR.}
    \label{fig:sensitivity-latency-breakdown}
\end{figure}

\smallskip
\noindent\textbf{Sensitivity Study.}
To validate the TM engine configuration
(Section~\ref{sec:token-merging-engine}), we evaluate offline encoder-only
merging and average results across Base, Small, and Medium. As shown in
Fig.~\ref{fig:sensitivity-latency-breakdown}(a), reducing the bitwidth of the matching Keys
from 16 to 4 decreases $\Delta$WER relative to Original from 3.42 to 0.38
percentage points (lower is better), while maintaining a merging ratio of
72.61\%. With matching thresholds held fixed, lower-bitwidth quantization may
make matching more conservative, leaving some important tokens unmerged.
At 2 bits, $\Delta$WER decreases
further, but the merging ratio falls to 29.10\%, suggesting a loss of similarity
information. These results support our 4-bit setting, which retains sufficient
similarity information with low hardware overhead.

In Fig.~\ref{fig:sensitivity-latency-breakdown}(b), increasing the active cache
from 1 to 2 entries lowers WER while achieving a slightly higher merging ratio.
With one entry, tokens can match only the most recent group. A second entry
keeps an earlier group available and can reduce missed matches between similar
tokens separated in the sequence. The lower $\Delta$WER suggests that this
grouping better preserves acoustic features. Larger caches offer little further
benefit but require more storage and may increase comparisons per token. Our
2-entry cache is therefore sufficient to retain useful recent groups with low
matching overhead.

\begin{figure}[!t]
    \centering
    \begingroup
        \scriptsize
        \renewcommand{\arraystretch}{1.15}
        \setlength{\tabcolsep}{1pt}
        \sbox{\areabreakdownbox}{%
        \begin{tabular*}{0.65\columnwidth}{@{\extracolsep{\fill}}lcc}
            \toprule
            \textbf{Component} & \textbf{Area (mm$^2$)} & \textbf{Area Ratio (\%)} \\
            \midrule
            Matrix Multiply Unit & 1.98 & 24.55 \\
            Act. \& Weight Buffers & 3.91 & 48.48 \\
            Vector Processing Unit & 0.66 & 8.22 \\
            \textbf{Token Merging Engine} & \textbf{0.33} & \textbf{4.05} \\
            Other & 1.19 & 14.70 \\
            \midrule
            Total & 8.06 & 100.00 \\
            \bottomrule
        \end{tabular*}%
        }
        \begin{tabular}{@{}c@{\hspace{0.01\columnwidth}}c@{}}
            \makebox[0.65\columnwidth]{\raisebox{\dp\areabreakdownbox}{\usebox{\areabreakdownbox}}} &
            \makebox[0.34\columnwidth]{\includegraphics[trim=0 0 0 2bp,clip,
                height=\dimexpr\ht\areabreakdownbox+\dp\areabreakdownbox\relax]{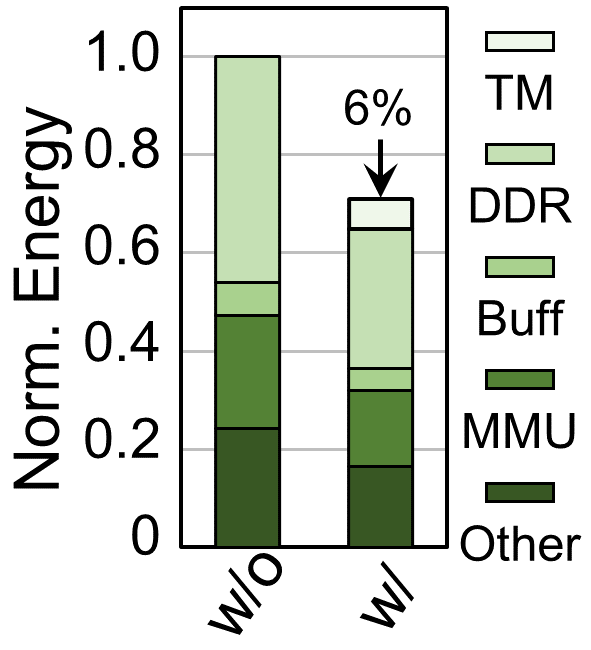}} \\[5pt]
            (a) Area breakdown & (b) Energy consumption
        \end{tabular}
    \endgroup
    \caption{(a) Area breakdown of the MergeASR accelerator. (b) Normalized
    energy consumption without (w/o) and with (w/) token merging during a
    single offline encoder inference for Whisper-Medium.}
    \label{fig:area-inference-energy}
\end{figure}

\smallskip
\noindent\textbf{Latency Breakdown.}
Fig.~\ref{fig:sensitivity-latency-breakdown}(c) profiles offline encoder
latency across Base, Small, and Medium. On the GPU with token merging, MLPs
process fewer tokens and take less time than attention. However, without
dedicated dataflow support, token merging dominates latency in the GPU
implementation profiled here. In contrast, MergeASR's dedicated TM engine
efficiently handles matching and overlaps it with attention operations, keeping
merging at only 2.23\% of total encoder latency on average. Relative to the
baseline NPU, MergeASR reduces MLP latency by 71.3\% and achieves an overall
encoder speedup of 1.38$\times$. Despite its much smaller compute
and memory bandwidth budgets, MergeASR achieves 45.3\% of the GPU's encoder
throughput.

\smallskip
\noindent\textbf{Area and Energy.}
Fig.~\ref{fig:area-inference-energy}(a) shows the area breakdown of the
MergeASR accelerator. The Compute Core occupies 81.3\% of
the area, making it costly to add more compute units. In contrast, the TM
engine occupies only 4.1\% yet enables an average encoder
latency reduction of 27.0\% relative to the baseline NPU.
Fig.~\ref{fig:area-inference-energy}(b) compares offline encoder energy with
and without token merging. With a token-merging energy overhead of only 6\%
relative to the baseline, MergeASR achieves an energy reduction of 29.0\%.
Encoder MLPs expand their inputs by 4$\times$, producing large intermediate
data stored in DDR. Token merging reduces the data volume and DDR traffic,
lowering energy consumption.

\section{Discussion and Limitations}
\label{sec:discussion}

\subsection{Effects of Component-Specific Merging}
\label{sec:discussion-component-specific}

Table~\ref{tab:component-policy-ablation} shows the effects of merging
operations and matching policies on inference performance and transcription
quality. MergeASR focuses on encoder MLPs because they have a more regular
dataflow and are a major source of encoding latency, while also supporting
token merging in self-attention. Merging MLPs provides a better trade-off:
1.37$\times$ speedup with a WER increase of 0.18 percentage points, compared
with 1.15$\times$ and 0.26 percentage points for self-attention. Merging both
operations with shared matching results within each layer achieves
1.86$\times$ speedup with a 0.60-point WER increase. Joint merging may make
token representations more similar in later layers, enabling further
merging. It therefore offers further speedup when a larger loss in
transcription quality is acceptable.

Unlike the uniform matching policy in prior
work~\cite{yoo2024adaptiv,bolya2022token,shin2025atom}, MergeASR performs
matching at each encoder layer and reuses first-layer matching results
across decoder layers, based on the acoustic-token similarity analysis in
Section~\ref{sec:acoustic-token-similarity}.
Table~\ref{tab:component-policy-ablation} shows that encoder reuse achieves
only 1.31--1.32$\times$ speedup, as the first layer's merging ratio limits
token reduction in later layers. The WER increase of 0.96--0.99 percentage
points is consistent with the observation that encoder similarity changes
across layers, making reused matching results less reliable. In contrast,
decoder similarity remains more stable across layers. First-layer reuse achieves similar
transcription quality to per-layer matching while avoiding repeated
matching, slightly increasing speedup from 1.48$\times$ to 1.49$\times$,
compared with 1.37$\times$ without decoder merging. These results support
the component-specific policy in Section~\ref{sec:merging-unmerging}.

\begin{table}[!t]
    \centering
    \caption{Effect of merging operations and matching policies.}
    \label{tab:component-policy-ablation}
    \small
    \setlength{\tabcolsep}{3pt}
    \renewcommand{\arraystretch}{1.15}
    \begin{tabular*}{\linewidth}{@{\extracolsep{\fill}}lllcc@{}}
        \toprule
        \multicolumn{2}{c}{\textbf{Encoder}} &
        \multicolumn{1}{c}{\textbf{Decoder}} &
        \multicolumn{1}{c}{\textbf{Speedup}} &
        \multicolumn{1}{c}{\textbf{$\Delta$WER}} \\
        \cmidrule(lr){1-2}\cmidrule(lr){3-3}
        \textbf{Operation} & \textbf{Matching} & \textbf{Matching} &
        ($\times$) & (pp) \\
        \midrule
        Off & \textemdash & Off & 1.00 & 0.00 \\
        \addlinespace[3pt]
        SA$^{\dagger}$ & Per-layer & Off & 1.15 & +0.26 \\
        SA$^{\dagger}$ + MLP & Per-layer & Off & 1.86 & +0.60 \\
        MLP & Per-layer & Off & 1.37 & +0.18 \\
        \midrule
        MLP & Reuse$^{\ddagger}$ & Per-layer & 1.31 & +0.96 \\
        MLP & Reuse$^{\ddagger}$ & Reuse$^{\ddagger}$ & 1.32 & +0.99 \\
        MLP & Per-layer & Per-layer & 1.48 & +0.32 \\
        \textbf{MLP} & \textbf{Per-layer} & \textbf{Reuse}$^{\ddagger}$ & \textbf{1.49} & \textbf{+0.29} \\
        \bottomrule
    \end{tabular*}
    \par\smallskip
    \noindent\footnotesize\raggedright
    $^{\dagger}$SA: self-attention.\quad $^{\ddagger}$Reuse: first-layer matching reuse.\par
    All results are for Whisper-Medium in streaming mode. Evaluation is performed on the MergeASR hardware simulator, with WER measured on LibriSpeech test-other.
\end{table}

\subsection{Threshold Selection}
\label{sec:discussion-limitations}

Some methods use the same similarity threshold across layers or directly
set the number of tokens to merge~\cite{bolya2022token,shin2025atom}.
However, layers may differ in token similarity and sensitivity to merging.
Some layers may require more
conservative merging to preserve acoustic details, while others contain
more redundancy. We empirically select thresholds for different
layers to balance the merging ratio and transcription quality. A systematic
selection policy based on layer characteristics and accuracy or
latency requirements could provide more direct control over this trade-off.
It could also reduce the effort needed to configure merging across models
and inference settings.

\section{Conclusion}
\label{sec:conclusion}

Modern ASR inference faces substantial computational demands from long
sequences of acoustic tokens. This paper presented MergeASR, a
hardware--software co-designed token-merging framework that identifies and
exploits acoustic-token redundancy online. It combines token merging
tailored to ASR with a lightweight hardware engine. By allowing similar acoustic tokens to share
computation, the framework accelerates both offline and streaming ASR inference. These
benefits make it a practical solution for efficient deployment of
modern ASR models and enable more intelligent voice interaction.

\section*{Acknowledgment}

GPT-6.1 Sol (OpenAI) was used to assist with language polishing and LaTeX
formatting. All content was reviewed and approved by the authors.

\bibliographystyle{IEEEtran}
\bibliography{refs}

\end{document}